\documentclass[a4paper,11pt]{article}

\usepackage{jcappub}
\usepackage{hyperref}
\usepackage{graphicx}

\usepackage[T1]{fontenc} 

\begin{document}

\title{Early-Matter-Like Dark Energy and the Cosmic Microwave Background}

\author{R.\ Aurich}
\author{and S.\ Lustig}

\emailAdd{ralf.aurich@uni-ulm.de}
\emailAdd{sven.lustig@uni-ulm.de}

\affiliation{Institut f\"ur Theoretische Physik, Universit\"at Ulm,\\
Albert-Einstein-Allee 11, D-89069 Ulm, Germany}

\abstract{
Early-matter-like dark energy is defined as a dark energy component
whose equation of state approaches that of cold dark matter
(CDM) at early times.
Such a component is an ingredient of unified dark matter (UDM) models,
which unify the cold dark matter and the cosmological constant of
the $\Lambda$CDM concordance model into a single dark fluid.
Power series expansions in conformal time of the perturbations of the
various components for a model with early-matter-like dark energy are provided.
They allow the calculation of the cosmic microwave background (CMB)
anisotropy from the primordial initial values of the perturbations.
For a phenomenological UDM model,
which agrees with the observations of the local Universe,
the CMB anisotropy is computed and compared with the CMB data.
It is found that a match to the CMB observations is possible
if the so-called effective velocity of sound $c_{\hbox{\scriptsize eff}}$
of the early-matter-like dark energy component is very close to zero.
The modifications on the CMB temperature and polarization power spectra
caused by varying the effective velocity of sound are studied.
}


\arxivnumber{1511.01691}

 
\maketitle
\flushbottom

\section{Introduction}
\label{sec:intro}

The $\Lambda$CDM concordance model successfully describes a
variety of cosmological observations over many orders of length scales.
However, only about five percent of the energy density of the
universe can be ascribed to known particles of the standard model
of elementary particle physics.
The cold dark matter (CDM) and the dark energy components of the $\Lambda$CDM
are introduced as two phenomenological components.
There have been many attempts to naturally explain the dark energy
in the framework of quintessence models,
which explain it in terms of a scalar field governed by some potential.
In contrast to the cosmological constant $\Lambda$ of the $\Lambda$CDM model,
which possesses a constant equation of state $w=p/\varepsilon=-1$
with the pressure $p$ and the energy density $\varepsilon$,
quintessence scenarios allow a time-varying equation of state.
Besides the equation of state, there are further degrees of freedom
which characterise a possible phenomenological component as emphasised
by \cite{Hu_1998}, who introduced a generalised dark matter component
defined by the effective velocity of sound $c_{\hbox{\scriptsize eff}}$
(in the rest frame of the dark component) as well as a viscosity velocity
$c_{\hbox{\scriptsize vis}}$ which is related to the anisotropic stress.
The usual cold dark matter is characterised by $w=0$,
$c^2_{\hbox{\scriptsize eff}}=0$ and $c^2_{\hbox{\scriptsize vis}}=0$,
whereas for quintessence models, the parameters are
$c^2_{\hbox{\scriptsize eff}}=1$ and $c^2_{\hbox{\scriptsize vis}}=0$
and $w$ is restricted to $-1\leq w\leq 1$.

Since the $\Lambda$CDM model requires two phenomenological components,
one could ponder the issue
whether a single phenomenological dark component might suffice.
This component could have a time-varying equation of state $w=w(t)$ and
possibly non-standard values of $c^2_{\hbox{\scriptsize eff}}$ and
$c^2_{\hbox{\scriptsize vis}}$.
In order to describe the accelerated expansion of the recent cosmos,
one needs $w(t)<-\frac 13$ for recent times.
On the other hand, one has to ensure that the structure formation
in the baryonic sector can evolve around the time of recombination
as described in the framework of the $\Lambda$CDM model,
which requires a significant dark matter component having
$w(t)=0$ at early times.
A phenomenological component with such a behaviour is called a
unified dark matter (UDM) component.

The prototypical model for a UDM component is the Chaplygin gas
which is originally considered in a cosmological context in
\cite{Kamenshchik_Moschella_Pasquier_2001},
see also
\cite{Fabris_Goncalves_deSouza__2001,Bilic_Tupper_Viollier_2002,%
Bento_Bertolami_Sen_2002,Carturan_Finelli_2003,Makler_deOliveira_Waga_2003}
for other early works and for the generalised Chaplygin gas
having an equation of state $p = -A/\varepsilon^\alpha$ with $0\leq \alpha \leq 1$.
In \cite{Sandvik_Tegmark_Zaldarriaga_2004}
it is shown that the parameter space of the generalised Chaplygin gas
is confined to a very small domain with $\alpha$ so close to zero
that it is nearly identical to the $\Lambda$CDM model
such that the motivation for the Chaplygin gas is undermined.
For a discussion of the case $\alpha>1$,
where a causality problem arises, see \cite{Piattella_2010}.
A loophole for the restricted parameter space is provided
by allowing the generation of entropy perturbations from
adiabatic initial conditions.
This corresponds to an effective speed of sound $c^2_{\hbox{\scriptsize eff}}$
different from the adiabatic one \cite{Reis_Waga_Calvao_Joras_2003}.
In that paper, it is shown that a vanishing effective speed of sound
$c^2_{\hbox{\scriptsize eff}}$ suffices to reconcile
the generalised Chaplygin gas with the matter power spectra $P(k)$
for a large range of the parameter $\alpha$.

Also models with a non-canonical kinetic Lagrangian,
so-called $k$-essence models,
can provide a scenario for an UDM model as it is put forward in
\cite{Scherrer_2004}
where the simplest class of $k$-essence models with a constant
potential term is considered.
It is emphasised in \cite{Scherrer_2004} that a low sound speed
in these models circumvents the difficulties of the Chaplygin gas
discussed above
since the integrated Sachs-Wolfe effect is then suppressed
at large-angular scales
\cite{Erickson_Caldwell_Steinhardt_Picon_Mukhanov_2002,DeDeo_Caldwell_Steinhardt_2003}.

For a review discussing various unified dark matter scenarios,
see \cite{Bertacca_Bartolo_Matarres_2010},
and see, for example, also
\cite{Hipolito_Velten_Zimdahl__2009,%
Piattella_Bertacc_Bruni_Pietrobon_2010,%
Bertacca_Bruni_Piattella_Pietrobon_2011,%
Campos_Fabris_Perez_Piattella_2013,%
Borges_Carneiro_Fabris_Zimdahl_2013,%
Bruni_Lazkoz_Fernandez_2013,%
Carames_Fabris_Velten_2014,%
Kumar_Sen_2014,%
Lazkoz_Leanizbarrutia_Salzano_2015}
for further details regarding unified dark matter models.

The above scenarios are based on a Lagrangian and thus
provide a physical justification of the models.
However, because of the vast number of possible models,
one can retreat to the pure phenomenological aspect.
In this respect, one considers the dark energy as
generalised dark matter in terms of the parameters
$c_{\hbox{\scriptsize eff}}$ and $c_{\hbox{\scriptsize vis}}$
as defined in \cite{Hu_1998}.
The UDM models require a dark matter component which has an equation of state
$w$ close to zero around the time of recombination.
In addition, we will assume here
that the equation of state will be almost zero deep in the radiation area.
This contrasts to some tracker scenarios where the dark energy component
behaves radiation-like in the radiation area.
This paper focuses on early-matter-like dark energy
which is defined by a dark energy component,
whose equation of state vanishes at early times
$w_{\hbox{\scriptsize de}}(\eta=0)=0$
such that it can be expanded as
\begin{equation}
\label{Def:Early-Matter-Like}
w_{\hbox{\scriptsize de}}(x) \; = \; w_1 x \; + \; w_2 x^2 \; + \; O(x^3)
\hspace{10pt} \hbox{ with } \hspace{10pt}
x := \frac{a(\eta)}{A_0}
\hspace{10pt} ,
\end{equation}
where $A_0$ denotes the  current value of the scale factor $a(\eta)$
so that $x\in[0,1]$.
It is the aim of this paper to provide power series of the
perturbations of the various components in terms of the
conformal time $\eta$.
These allow the computation of the initial conditions for the numerical
integration of the perturbations shortly before the recombination
such that a tight-coupling approximation is not necessary.
Although this Introduction emphasises UDM models,
the power series is also computed for the usual cold dark matter component.
Thus, the equations can also be applied to models dealing with
dark matter and dark energy as two separated components
provided that the equation of state of dark energy is given by
(\ref{Def:Early-Matter-Like}).
The algorithm is then applied to a phenomenological unified model
which is suggested in \cite{Cuzinatto_Medeiros_deMorais_2014}
where the equation of state
\begin{equation}
\label{Def:model_eos}
w(z) \; = \; \frac 1\pi \arctan(\alpha z - \beta) - \frac 12
\end{equation}
is assumed with two parameters $\alpha$ and $\beta$.
It is shown in \cite{Cuzinatto_Medeiros_deMorais_2014} that such a
phenomenological dark fluid can describe the supernovae Ia data,
$\gamma$-ray bursts and the baryon acoustic oscillations for a parameter set
around $\alpha\simeq 2.14$ and $\beta\simeq 0.95$.
Since this phenomenological dark fluid can describe these data,
the question emerges whether this model predicts a power spectrum of
the cosmic microwave background radiation according to
cosmological observations.

\section{The background model}

In order to derive the power series of the perturbations,
the power series of the scale factor $a(\eta)$ of the background model
has to be computed at first.
Since the early-matter-like dark energy is important at early times,
in contrast to models with a dark energy with $w(\eta)<0$,
the computation of the scale factor $a(\eta)$ has to take
the dark energy component into account.
The energy density $\varepsilon_{\hbox{\scriptsize de}}(\eta)$ evolves as
\begin{eqnarray}\nonumber 
\\ \label{Eq:Def_epsilon_de}
\varepsilon_{\hbox{\scriptsize de}}(\eta) & = &
\varepsilon_{\hbox{\scriptsize de}}^0 \, \exp\left( 3 \int_{x(\eta)}^1
\frac{1+w_{\hbox{\scriptsize de}}(x)}x\, dx \right)
\\ & = & \nonumber 
\varepsilon_{\hbox{\scriptsize de}}^0 \, \exp\left( 3 \int_0^1
\frac{w_{\hbox{\scriptsize de}}(x)}x\, dx \right)
\, \exp\left( 3 \int_{x(\eta)}^1 \frac{dx}x \right)
\\ & & \nonumber 
\hspace{130pt} \times
\; \exp\left(-3 \int_0^{x(\eta)} \frac{w_{\hbox{\scriptsize de}}(x)}x\, dx \right)
\\ & = & \nonumber 
\varepsilon_{\hbox{\scriptsize de}}^{\hbox{\scriptsize eff}} \,
\left(\frac{A_0}{a(\eta)}\right)^3
\, \exp\left(-3 \int_0^{x(\eta)} \frac{w_{\hbox{\scriptsize de}}(x)}x\, dx \right)
\hspace{10pt} ,
\end{eqnarray}
where $\varepsilon_{\hbox{\scriptsize de}}^0$ is the current energy density and we have defined
\begin{equation}
\label{Def:epsilon_de_eff}
\varepsilon_{\hbox{\scriptsize de}}^{\hbox{\scriptsize eff}} \; := \;
\varepsilon_{\hbox{\scriptsize de}}^0 \, \exp\left( 3 \int_0^1
\frac{w_{\hbox{\scriptsize de}}(x)}x\, dx \right)
\hspace{10pt} .
\end{equation}
Note that in eqs.\,(\ref{Eq:Def_epsilon_de}) and (\ref{Def:epsilon_de_eff}),
the full expression of the equation of state has to be used and not only
the early-times approximation (\ref{Def:Early-Matter-Like}).
Furthermore, it is convenient to define
\begin{equation}
\label{Def:Omega_de_eff}
\Omega_{\hbox{\scriptsize de}}^{\hbox{\scriptsize eff}} \; := \;
\frac{8\pi G}{3 H_0^2c^2} \, \varepsilon_{\hbox{\scriptsize de}}^{\hbox{\scriptsize eff}}
\end{equation}
with the Hubble constant $H_0$ and the gravitational constant $G$.
The quantity $\Omega_{\hbox{\scriptsize de}}^{\hbox{\scriptsize eff}}$
can be interpreted as the effective matter contribution of the UDM component
at early times and should not be confused with the current value
of the dark energy density
$\Omega_{\hbox{\scriptsize de}}=\frac{8\pi G}{3 H_0^2c^2} \varepsilon_{\hbox{\scriptsize de}}^0$.

The Friedmann equation leads to the expansion of the scale factor $a(\eta)$
in terms of the conformal time $\eta$
\begin{equation}
\label{Eq:scale_factor}
a(\eta) \; = \; a_1 \, \eta \, + \, a_2 \, \eta^2 \, + \, a_3 \, \eta^3 \, + \, O(\eta^4)
\end{equation}
with the coefficients
\begin{equation}
\label{Eq:scale_factor_coeff}
a_1 \; = \; \frac{H_0}{c} \, A_0^2\, \sqrt{\Omega_{\hbox{\scriptsize rad}}}
\hspace{10pt} , \hspace{10pt}
a_2 \; = \; \frac 14 \left(\frac{H_0}{c}\right)^2 A_0^3 \,
(\Omega_{\hbox{\scriptsize mat}}+\Omega_{\hbox{\scriptsize de}}^{\hbox{\scriptsize eff}}\,)
\end{equation}
and
\begin{equation}
\label{Eq:scale_factor_coeff_3}
a_3 \; = \; \frac 16 \left(\frac{H_0}{c}\right)^3 A_0^4 \,
\sqrt{\Omega_{\hbox{\scriptsize rad}}} \, \left(\Omega_{\hbox{\scriptsize curv}}-
3\, w_1\, \Omega_{\hbox{\scriptsize de}}^{\hbox{\scriptsize eff}}\,\right)
\end{equation}
with the usual definitions of the present-day densities
$\Omega_{\hbox{\scriptsize rad}}=\Omega_\gamma+\Omega_\nu$,
$\Omega_{\hbox{\scriptsize mat}}=\Omega_{\hbox{\scriptsize bar}}+\Omega_{\hbox{\scriptsize cdm}}$
and $\Omega_{\hbox{\scriptsize curv}}$.
In the case of UDM models, one has $\Omega_{\hbox{\scriptsize cdm}}=0$.
Equation (\ref{Eq:scale_factor_coeff}) shows that
$\Omega_{\hbox{\scriptsize mat}}$ and $\Omega_{\hbox{\scriptsize de}}^{\hbox{\scriptsize eff}}$
contribute in the same manner to the early-times behaviour of the scale factor.
This expansion takes the early-matter-like dark energy into account
and is used for the derivation of the power expansions of the
perturbations of the various components.

\section{Initial conditions and evolution of perturbations}

The description of the perturbations requires the definition of the gauge.
In this paper, we use the conformal Newtonian gauge,
where the metric takes the form
\begin{equation}
\label{Eq:Metric}
ds^2 \; = \;
a^2(\eta) \Big\{ -(1+2\Psi) d\eta^2 \, + \,
(1+2\Phi) \gamma_{ij} dx^i dx^j \Big\}
\end{equation}
and $\gamma_{ij}$ denotes the spatial metric.
The Newtonian metric perturbations are characterised by the
two scalar potentials $\Psi$ and $\Phi$.
Note that the sign of $\Phi$ differs from the definition used
in \cite{Ma_Bertschinger_1995} but is identical to the choice
of \cite{Hu_1995}.

In this paper, only isentropic initial conditions are considered
which lead to initial perturbations in the relative energy density
perturbations in dependence on the equation of state $w_x$ at $\eta=0$
and on the initial perturbation of the potential $\Psi_0$
\begin{equation}
\label{Eq:initial_condition}
\delta_{x,0} \; = \; -\frac 32\, (1+w_x) \, \Psi_0
\hspace{10pt} .
\end{equation}
These conditions are the natural outcome,
if the primordial perturbations are generated from a single degree of freedom.
This is the case in many inflationary scenarios where the decay of the
inflaton leads to the stated isentropic initial conditions.
In Appendix \ref{app:isentropic_initial_condition}, it is demonstrated
that the initial conditions (\ref{Eq:initial_condition}) lead to pure
isentropic initial conditions without any isocurvature contribution.

The dynamical evolution of the perturbation of the dark component is not
entirely fixed by specifying the initial condition (\ref{Eq:initial_condition})
and the equation of state $w_{\hbox{\scriptsize g}}$,
since this does not determine the pressure perturbation
$\delta p_{\hbox{\scriptsize g}}$ which is a priori unknown in this case.
The equation of state
$p_{\hbox{\scriptsize g}}=w_{\hbox{\scriptsize g}}\varepsilon_{\hbox{\scriptsize g}}$
of the generalised dark matter component determines the
adiabatic speed of sound as
\begin{equation}
\label{Eq:sos_adiabatic}
c_g^2 \; = \;
\frac{\dot p_{\hbox{\scriptsize g}}}{\dot\varepsilon_{\hbox{\scriptsize g}}}
\; = \;
w_{\hbox{\scriptsize g}} \, - \, 
\frac 13\, \frac{\dot w_{\hbox{\scriptsize g}}}{1+w_{\hbox{\scriptsize g}}}\,
\frac{a}{\dot a}
\hspace{10pt} ,
\end{equation}
which enters the evolution equations.
However, the entropy perturbation $\Gamma$ is not fixed by specifying $w_g$.
But this is necessary, since this quantity also occurs in the
evolution equations of the density perturbation \cite{Hu_1998}
\begin{equation}
\label{Eq:gdm_density}
\dot\delta_{\hbox{\scriptsize g}} \; = \;
-\, (1+w_{\hbox{\scriptsize g}})(kv_{\hbox{\scriptsize g}}+3\dot\Phi)
\, + \,
\frac{\dot w_{\hbox{\scriptsize g}}}{1+w_{\hbox{\scriptsize g}}}\,
\delta_{\hbox{\scriptsize g}}
\, - \,
3\, \frac{\dot a}a\,
w_{\hbox{\scriptsize g}} \, \Gamma
\end{equation}
and of the velocity
\begin{equation}
\label{Eq:gdm_velocity}
\dot v_{\hbox{\scriptsize g}} \; = \;
-\, \frac{\dot a}a \,(1-3c^2_{\hbox{\scriptsize g}})\,v_{\hbox{\scriptsize g}}
\, + \,
\frac{c_{\hbox{\scriptsize g}}^2}{1+w_{\hbox{\scriptsize g}}}
k \delta_{\hbox{\scriptsize g}}
\, + \,
\frac{k w_{\hbox{\scriptsize g}}\Gamma}{1+w_{\hbox{\scriptsize g}}}
\, + \, k \Psi
\hspace{10pt} ,
\end{equation}
where the scalar anisotropic stress amplitude is assumed to be zero.
The entropy perturbation $\Gamma$ occurring in
eqs.\,(\ref{Eq:gdm_density}) and (\ref{Eq:gdm_velocity})
is related to the freedom of specifying the
pressure perturbation \cite{Mukhanov_Feldman_Brandenberger_1992,Hu_1998}.
It is convenient to parameterise the entropy perturbation $\Gamma$
by the effective speed of sound $c^2_{\hbox{\scriptsize eff}}$ as
\begin{equation}
\label{Eq:gdm_etropy_perturbation}
\Gamma \; = \;
\frac{c^2_{\hbox{\scriptsize eff}}-c^2_{\hbox{\scriptsize g}}}{w_{\hbox{\scriptsize g}}}
\,\delta_{\hbox{\scriptsize g}}^{\hbox{\scriptsize (rest)}}
\hspace{10pt} ,
\end{equation}
where the density perturbation
$\delta_{\hbox{\scriptsize g}}^{\hbox{\scriptsize (rest)}}$
in the rest frame of the
generalised dark matter is determined by the transformation
\begin{equation}
\label{Def:d_rest}
\delta_{\hbox{\scriptsize g}}^{\hbox{\scriptsize (rest)}} \; = \;
\delta_{\hbox{\scriptsize g}} \, + \,
3 \frac{\dot a}a\, (1+w_{\hbox{\scriptsize g}}) \, \frac{v_{\hbox{\scriptsize g}}}k
\hspace{10pt} .
\end{equation}
If the effective speed of sound
$c^2_{\hbox{\scriptsize eff}}$ coincides with the adiabatic one
$c^2_{\hbox{\scriptsize g}}$, the entropy perturbation $\Gamma$ vanishes.
The effective speed of sound $c^2_{\hbox{\scriptsize eff}}$ is considered
as a free parameter defining the evolution of the perturbations
of the dark fluid.
The effective speed of sound can be interpreted as the rest frame sound speed
\begin{equation}
c^2_{\hbox{\scriptsize eff}} \; = \;
\frac{\delta p_g^{\hbox{\scriptsize (rest)}}}
{\delta \rho_g^{\hbox{\scriptsize (rest)}}}
\hspace{10pt} ,
\end{equation}
where $\delta p_g^{\hbox{\scriptsize (rest)}}$ and
$\delta \rho_g^{\hbox{\scriptsize (rest)}}$
denote the pressure and density perturbations in the rest frame
of the generalised dark matter component.
This leads to a definition independent of the chosen gauge.

In order to compute the CMB anisotropy, one has to integrate a coupled
system of differential equations.
The numerical integration cannot start at $\eta=0$,
where the isentropic initial conditions (\ref{Eq:initial_condition})
are posed, but only at a later time $\eta_{\hbox{\scriptsize start}}$
before the recombination takes place at $\eta_{\hbox{\scriptsize rec}}$
that is
$0 \ll \eta_{\hbox{\scriptsize start}} \ll \eta_{\hbox{\scriptsize rec}}$.
To ensure that the isentropic initial conditions (\ref{Eq:initial_condition})
defined at $\eta=0$ are correctly taken into account
at $\eta_{\hbox{\scriptsize start}}$,
a power series expansion for each of the perturbations is used.
Since the system of differential equations is coupled,
there is no direct way of obtaining such power series.
Instead one has to carry out a step by step calculation to
eliminate the unwanted quantities.
The strategy of finding the appropriate solutions is to express
the expansion coefficients of the potentials $\Psi(\eta)$ and $\Phi(\eta)$
solely in terms of the background model
that is in terms of the coefficients given in
(\ref{Eq:scale_factor_coeff}) and (\ref{Eq:scale_factor_coeff_3}).
These potential expansions are given in Appendix \ref{app:potentials}.
Then, the second step is to express the power series of the perturbations of
the various energy components solely in terms of the
expansion coefficients of the potentials $\Psi(\eta)$ and $\Phi(\eta)$.
In this way, the start values at $\eta_{\hbox{\scriptsize start}}$ are obtained
by computing at first the coefficients of the potentials
$\Psi(\eta)$ and $\Phi(\eta)$ given in Appendix \ref{app:potentials}
and thereafter, from the known potentials,
the perturbations $\delta_x(\eta_{\hbox{\scriptsize start}})$ and
$v_x(\eta_{\hbox{\scriptsize start}})$.
Their power series are given in Appendix \ref{app:matter_perturbation} and
Appendix \ref{app:radiation_perturbation} in terms of the potentials.
After all start values are computed at $\eta_{\hbox{\scriptsize start}}$
from the primordial initial values at $\eta=0$,
the numerical integration is carried out with the evolution equations
which are also given in Appendices \ref{app:potentials},
\ref{app:matter_perturbation} and \ref{app:radiation_perturbation}.
The metric and energy density perturbations allow the computation of the
CMB anisotropy along the lines given in \cite{Ma_Bertschinger_1995,Hu_1995}.
In contrast to public domain programs for the computation of the
CMB power spectrum,
our code uses the conformal Newtonian gauge instead of the synchronous gauge.

It should be noted that in \cite{Ballesteros_Lesgourgues_2010}
the series expansion of the potentials is derived by transforming the
corresponding result obtained in the synchronous gauge into the
conformal Newtonian gauge.
However, their result differs from that in Appendix \ref{app:potentials}
since the gauge transformation requires the synchronous gauge result
two order higher than used in \cite{Ballesteros_Lesgourgues_2010}.

\section{CMB anisotropy for the arctan-UDM model}
\label{CMB anisotropy}

In this section, we apply our program to the phenomenological UDM model
proposed in \cite{Cuzinatto_Medeiros_deMorais_2014} in order
to compute the CMB power spectrum.
The equation of state of this UDM component is already defined
in eq.\,(\ref{Def:model_eos}).
The parameters in the equation of state (\ref{Def:model_eos}) are fixed at
$\alpha= 2.14$ and $\beta= 0.95$,
since these values are in accordance with the supernovae Ia data,
$\gamma$-ray bursts and the baryon acoustic oscillations
as demonstrated in \cite{Cuzinatto_Medeiros_deMorais_2014}.
The equation of state (\ref{Def:model_eos}) leads
for the expansion (\ref{Def:Early-Matter-Like}) to the coefficients
\begin{equation}
\label{Eq:Coeff_arctan}
w_1 \; = \; - \, \frac 1{\pi\,\alpha}
\hspace{10pt} \hbox{ and } \hspace{10pt}
w_2 \; = \; - \, \frac{\alpha+\beta}{\pi\,\alpha^2}
\hspace{10pt} ,
\end{equation}
which have to be used for the computation of the start values
at $\eta_{\hbox{\scriptsize start}}$ as discussed above.

\begin{figure}
\begin{center}
\begin{minipage}{9cm}
\vspace*{-10pt}
\hspace*{-20pt}\includegraphics[width=9.0cm]{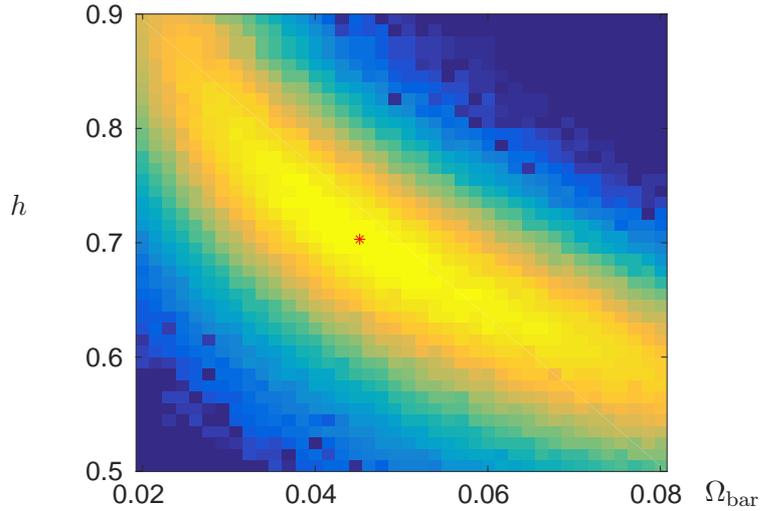}
\put(-10,12){$\Omega_{\hbox{\scriptsize bar}}$}
\put(-270,120){$h$}
\end{minipage}
\vspace*{-25pt}
\end{center}
\caption{\label{Fig:Probability_Obar_h}
The likelihood is plotted in the
$\Omega_{\hbox{\scriptsize bar}}-h$ plane.
The best-fit model is marked by the star.
}
\end{figure}

\begin{figure}
\begin{center}
\begin{minipage}{9cm}
\vspace*{-10pt}
\hspace*{-20pt}\includegraphics[width=9.0cm]{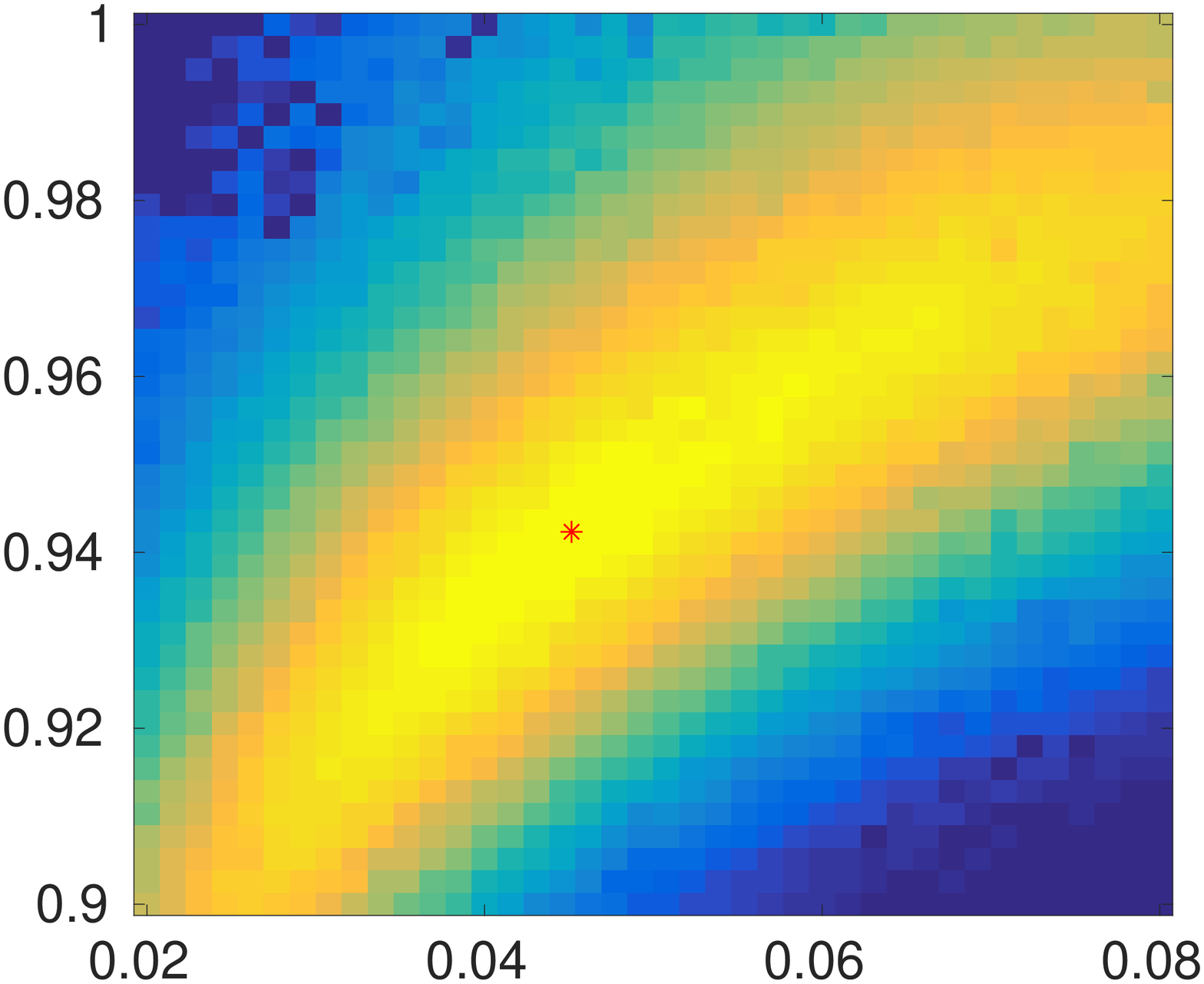}
\put(-10,12){$\Omega_{\hbox{\scriptsize bar}}$}
\put(-270,120){$\Omega_{\hbox{\scriptsize de}}$}
\end{minipage}
\vspace*{-25pt}
\end{center}
\caption{\label{Fig:Probability_Obar_Oq}
The likelihood is plotted in the
$\Omega_{\hbox{\scriptsize bar}}-\Omega_{\hbox{\scriptsize de}}$ plane.
The best-fit model is marked by the star.
}
\end{figure}

\begin{figure}
\begin{center}
\begin{minipage}{9cm}
\vspace*{-10pt}
\hspace*{-20pt}\includegraphics[width=9.0cm]{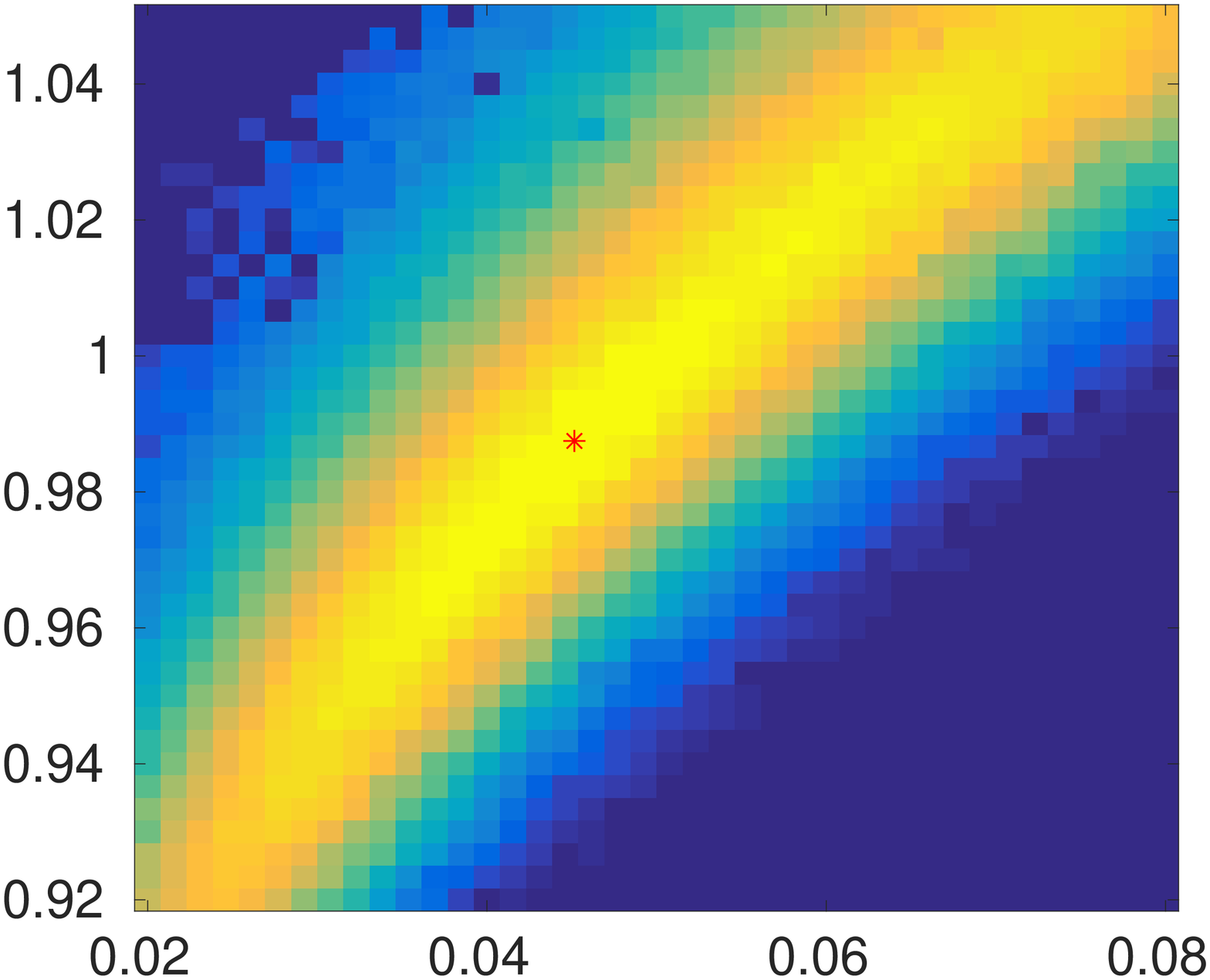}
\put(-10,12){$\Omega_{\hbox{\scriptsize bar}}$}
\put(-270,120){$\Omega_{\hbox{\scriptsize tot}}$}
\end{minipage}
\vspace*{-25pt}
\end{center}
\caption{\label{Fig:Probability_Obar_Otot}
The likelihood is plotted in the
$\Omega_{\hbox{\scriptsize bar}}-\Omega_{\hbox{\scriptsize tot}}$ plane.
The best-fit model is marked by the star.
}
\end{figure}

\begin{figure}
\begin{center}
\begin{minipage}{12cm}
\vspace*{-25pt}
\hspace*{-20pt}\includegraphics[width=12.0cm]{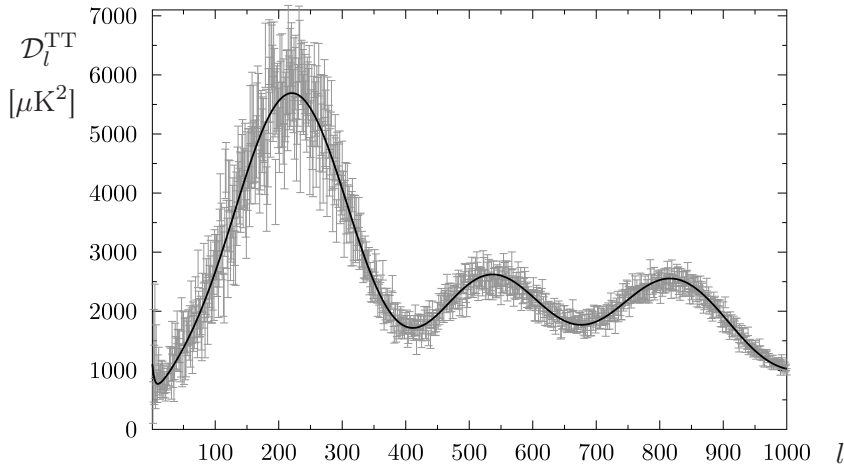}
\put(-34,22){$l$}
\put(-339,175){${\cal D}_l^{\hbox{\scriptsize TT}}$}
\put(-344,155){[$\mu\hbox{K}^2$]}
\end{minipage}
\vspace*{-25pt}
\end{center}
\caption{\label{Fig:Tl_best_fit_planck}
The angular power spectrum ${\cal D}_l^{\hbox{\scriptsize TT}} =
l(l+1) C_l^{\hbox{\scriptsize TT}} /(2\pi)$ is plotted
using the values of the best-fit model as stated in the text.
The ${\cal D}_l^{\hbox{\scriptsize TT}}$ measurements of the Planck 2015 data release
\cite{Planck_2015_I} are plotted with the corresponding error bars.
}
\end{figure}

\begin{figure}
\begin{center}
\begin{minipage}{12cm}
\vspace*{-25pt}
\hspace*{-20pt}\includegraphics[width=12.0cm]{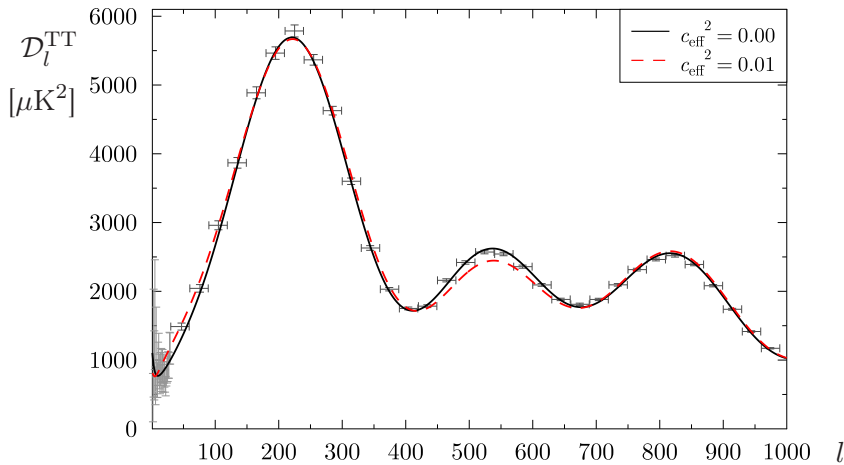}
\put(-34,22){$l$}
\put(-339,175){${\cal D}_l^{\hbox{\scriptsize TT}}$}
\put(-344,155){[$\mu\hbox{K}^2$]}
\end{minipage}
\vspace*{-25pt}
\end{center}
\caption{\label{Fig:Tl_best_fit_planck_binned}
The angular power spectrum ${\cal D}_l^{\hbox{\scriptsize TT}}$ is shown
for the best-fit model as stated in the text.
The binned data of the Planck 2015 data release
\cite{Planck_2015_I} are plotted for $l>30$
with the corresponding error bars.
In addition, the angular power spectrum is shown for a model
where in contrast to the best-fit model the effective speed of sound
is slightly enhanced to $c^2_{\hbox{\scriptsize eff}}=0.01$.
This leads to a reduced second acoustic peak due to a different normalisation
and spectral index.
}
\end{figure}

With the given values of $\alpha$ and $\beta$,
a Markov chain Monte Carlo (MCMC) sequence of models is generated
where $\Omega_{\hbox{\scriptsize bar}}$, $\Omega_{\hbox{\scriptsize de}}$
and the reduced Hubble constant $h$ are varied.
The other cosmological parameters are held fixed at values taken
from the $\Lambda$CDM concordance model.
The $\Lambda$CDM model with the parameters given in Table 9
in \cite{Planck_2015_I} designated as
``Planck TT+lowP+lensing'' is used in this paper.
The cosmological parameters of the $\Lambda$CDM model are
$\Omega_{\hbox{\scriptsize bar}}= 0.0484$,
$\Omega_{\hbox{\scriptsize cdm}}= 0.258$,
$\Omega_\Lambda= 0.6935$,
the reduced Hubble constant $h=0.678$,
the reionisation optical depth $\tau = 0.066$,
and the scalar spectral index $n_s=0.9677$.
In our UDM model, the densities
$\Omega_{\hbox{\scriptsize cdm}}$ and $\Omega_\Lambda$
are set to zero since they are unified in the component denoted by
$\Omega_{\hbox{\scriptsize de}}$.
For the reionisation optical depth, the value $\tau = 0.066$ is used.
Furthermore, the scalar spectral index $n_s$ and the overall
normalisation factor are fitted
so that a best-fit is obtained to the angular power spectrum
${\cal D}_l^{\hbox{\scriptsize TT}} := l(l+1) C_l^{\hbox{\scriptsize TT}}/(2\pi)$
of the Planck 2015 data release \cite{Planck_2015_I}.
The UDM model power spectra ${\cal D}_l^{\hbox{\scriptsize TT}}$ are fitted over 
the interval $2\leq l \leq 1000$ to the Planck data.

Furthermore, for this MCMC sequence, the value of the effective
speed of sound $c^2_{\hbox{\scriptsize eff}}$ is set to zero.
It is discussed below that non-zero values lead to the same problems
as in other UDM models as, for example, the generalised Chaplygin gas.
The likelihood distribution is shown in figures \ref{Fig:Probability_Obar_h}
to  \ref{Fig:Probability_Obar_Otot}
where the reduced Hubble constant $h$, the density of the dark energy
$\Omega_{\hbox{\scriptsize de}}$, and the total energy density
$\Omega_{\hbox{\scriptsize tot}}$ are shown in dependence on
$\Omega_{\hbox{\scriptsize bar}}$, respectively.
The best-fit model of this MCMC sequence has the cosmological parameters
$h = 0.703$,
$\Omega_{\hbox{\scriptsize bar}} = 0.0451$,
$\Omega_{\hbox{\scriptsize de}} = 0.942$,
$n_s = 0.975$.
These parameters lead to an effective early dark matter contribution of
$\Omega_{\hbox{\scriptsize de}}^{\hbox{\scriptsize eff}}=0.2267$
which is defined in (\ref{Def:Omega_de_eff}).
The best-fit model is marked by a star in figures \ref{Fig:Probability_Obar_h}
to  \ref{Fig:Probability_Obar_Otot}.
The corresponding angular power spectrum ${\cal D}_l^{\hbox{\scriptsize TT}}$ is shown in
figure \ref{Fig:Tl_best_fit_planck} in comparison to the unbinned Planck data
\cite{Planck_2015_I}.
The good agreement with the data is clearly visible.
A comparison with the binned Planck data is presented in figure
\ref{Fig:Tl_best_fit_planck_binned} which again reveals the agreement.
This plot also demonstrates the influence of a dark matter component with a
non-vanishing effective speed of sound,
where the very small value $c^2_{\hbox{\scriptsize eff}}=0.01$ is used.

\begin{figure}
\begin{center}
\begin{minipage}{12cm}
\vspace*{-25pt}
\hspace*{-20pt}\includegraphics[width=12.0cm]{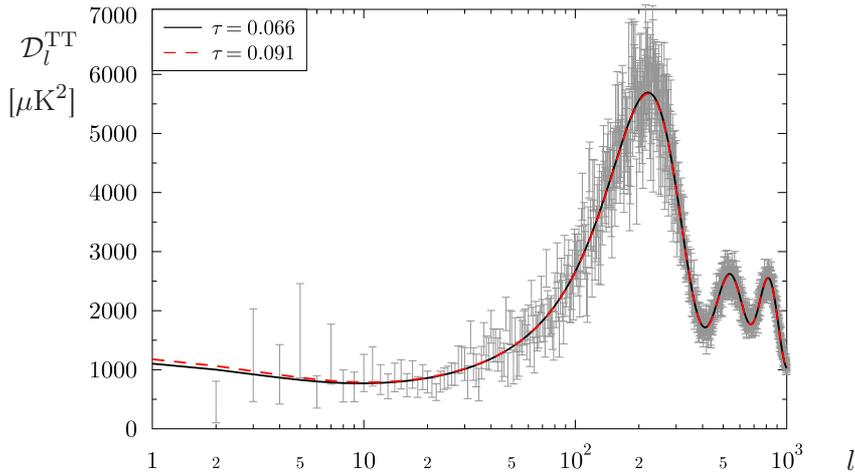}
\put(-30,19){$l$}
\put(-339,175){${\cal D}_l^{\hbox{\scriptsize TT}}$}
\put(-344,155){[$\mu\hbox{K}^2$]}
\end{minipage}
\vspace*{-25pt}
\end{center}
\caption{\label{Fig:Tl_best_fit_planck_log_tau}
The angular power spectrum ${\cal D}_l^{\hbox{\scriptsize TT}}$ is plotted
for the same model as in figure \ref{Fig:Tl_best_fit_planck}
but now with a logarithmic scaling on the $l$ axis
so that the agreement at low values of $l$ can be seen.
In addition, the best-fit model for the higher optical depth $\tau=0.091$
is shown as a dashed curve.
}
\end{figure}

The optical depth $\tau$ to the surface of last scattering has changed
from $\tau=0.089\pm 0.032$ of the Planck 2013 analysis \cite{Planck_Cosmo_Parameters_2013}
to $\tau=0.066\pm 0.016$ of the 2015 analysis \cite{Planck_2015_I}.
In order to test the influence of this change,
we also generate a MCMC sequence of models,
where the optical depth $\tau$ is fixed by the higher value of $\tau=0.091$.
It turns out, that this change does not alter the results,
since the best model possesses almost the same cosmological parameters
as those of the $\tau=0.066$ run.
Both model curves are shown in figure \ref{Fig:Tl_best_fit_planck_log_tau},
where a logarithmic scaling in $l$ is chosen
since the main difference occurs at low values of $l$.
It is seen that the curves are almost indistinguishable.

\begin{figure}
\begin{center}
\begin{minipage}{12cm}
\vspace*{-25pt}
\hspace*{-20pt}\includegraphics[width=12.0cm]{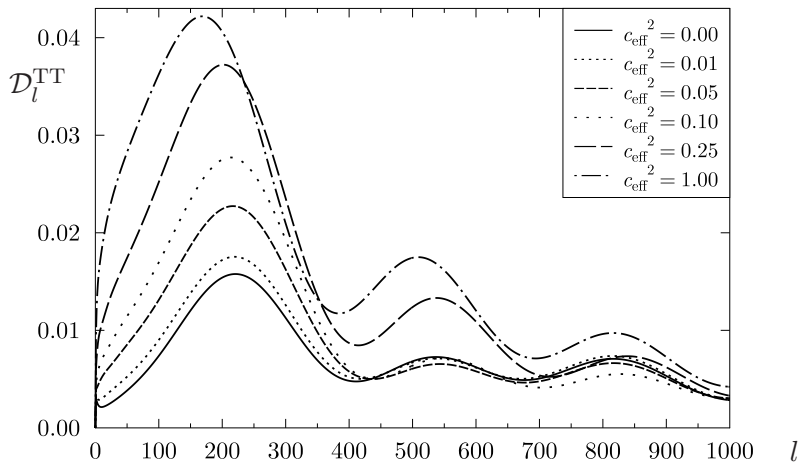}
\put(-30,22){$l$}
\put(-322,161){${\cal D}_l^{\hbox{\scriptsize TT}}$}
\end{minipage}
\vspace*{-25pt}
\end{center}
\caption{\label{Fig:Tl_as_fct_of_c_eff}
The angular power spectrum ${\cal D}_l^{\hbox{\scriptsize TT}}$ is shown in
dependence on the effective speed of sound $c^2_{\hbox{\scriptsize eff}}$.
All other cosmological parameters are kept constant with the values
of the best-fit model.
The same initial power power spectrum is used for all curves
and no fit to the cosmological data is performed.
}
\end{figure}

\begin{figure}
\begin{center}
\begin{minipage}{12cm}
\vspace*{-25pt}
\hspace*{-20pt}\includegraphics[width=12.0cm]{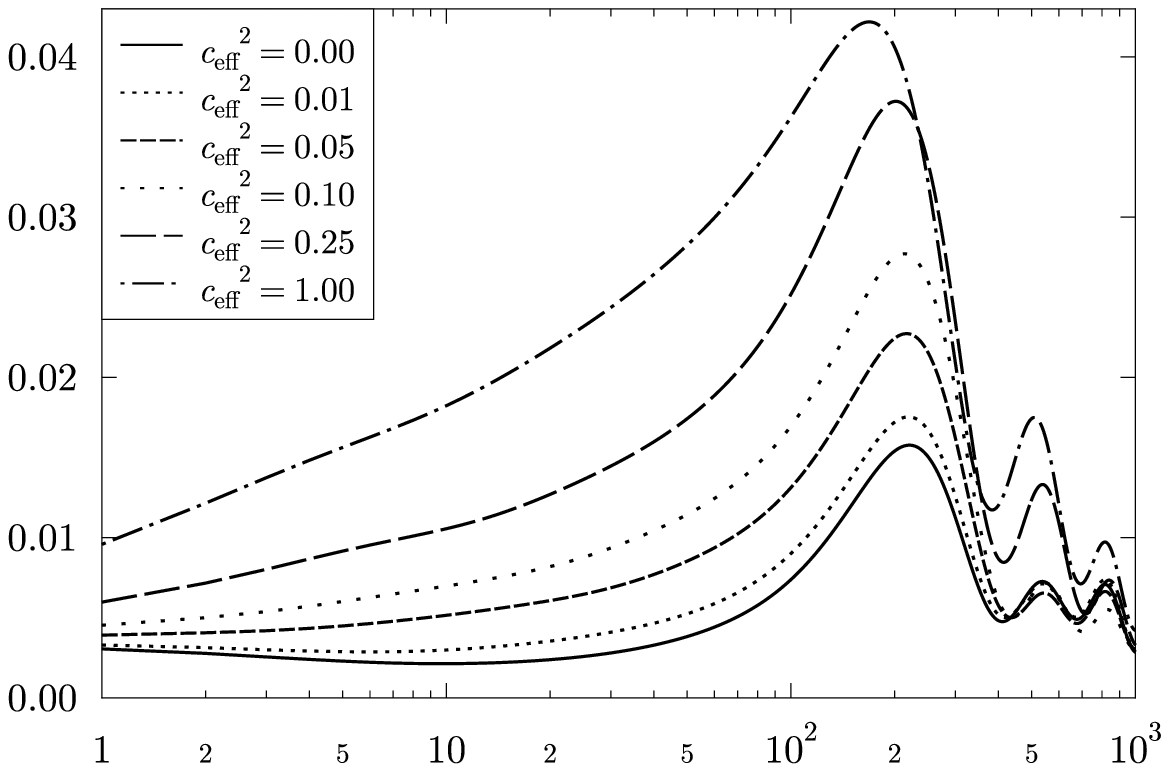}
\put(-30,19){$l$}
\put(-322,161){${\cal D}_l^{\hbox{\scriptsize TT}}$}
\end{minipage}
\vspace*{-25pt}
\end{center}
\caption{\label{Fig:Tl_as_fct_of_c_eff_log}
The same curves as in figure \ref{Fig:Tl_as_fct_of_c_eff} are shown
but now with a logarithmic scaling on the $l$ axis in order to emphasise
the late ISW contribution.
}
\end{figure}

The analysis of \cite{Cuzinatto_Medeiros_deMorais_2014} for the UDM model
does not require assumptions on the effective speed of sound $c^2_{\hbox{\scriptsize eff}}$.
Their analysis tests the agreement with the supernovae Ia data,
the baryon acoustic oscillations and the gamma-ray data,
and these tests are based on the background model providing the
luminosity distance and the angular diameter distance.
So it suffice to specify the equation of state (\ref{Def:model_eos})
for the UDM component.
The comparison with the CMB data requires the additional specification of
the effective speed of sound $c^2_{\hbox{\scriptsize eff}}$.
As discussed in the Introduction, other UDM models,
especially the generalised Chaplygin gas model,
have difficulties to match the behaviour of cosmological perturbations
in the case of a non-vanishing effective speed of sound.
In our main MCMC computation, we set $c^2_{\hbox{\scriptsize eff}}=0$
as discussed above.
However, also a shorter MCMC sequence is generated without this restriction
and reveals the preference for almost zero values of
$c^2_{\hbox{\scriptsize eff}}$.
Figure \ref{Fig:Tl_best_fit_planck_binned} shows a comparison of two models
which differ only in the effective speed of sound $c^2_{\hbox{\scriptsize eff}}$.
The values $c^2_{\hbox{\scriptsize eff}}=0$ and $c^2_{\hbox{\scriptsize eff}}=0.01$
are shown, and one observes a decline of the second acoustic peak
and an increasing amplitude towards small values of $l$
for the curve with $c^2_{\hbox{\scriptsize eff}}=0.01$.
However, these changes are mainly caused by fitting the model curve to the
CMB data.
Because of the increase of the amplitude at and prior to
the first acoustic peak,
one obtains a smaller overall normalisation factor and
a higher spectral index $n_s$,
which leads to a suppressed amplitude of the second peak
and an almost invariant amplitude of the third peak
for the two values of $c^2_{\hbox{\scriptsize eff}}$.

In order to avoid the influence of the normalisation factor
and the spectral index $n_s$,
we show in figures \ref{Fig:Tl_as_fct_of_c_eff} and
\ref{Fig:Tl_as_fct_of_c_eff_log} the angular power spectrum ${\cal D}_l^{\hbox{\scriptsize TT}}$
which is not fitted to the data.
These power spectra are computed from the same primordial power spectrum,
so that the genuine effect of the effective speed of sound
$c^2_{\hbox{\scriptsize eff}}$ can be seen.
All other cosmological parameters are those of the stated
best-fit model.
These plots reveal the large increase at small values of $l$
due to the integrated Sachs-Wolfe effect
\cite{Scherrer_2004,Bertacca_Bartolo_Matarres_2010}.
This is clearly visible in figure \ref{Fig:Tl_as_fct_of_c_eff_log},
where a logarithmic scaling is chosen for $l$.
Also the first acoustic peak increases significantly,
the amplitude of the second and third peak are relatively insensitive
for not too large values of $c^2_{\hbox{\scriptsize eff}}$.
For $c^2_{\hbox{\scriptsize eff}}=1$, which would be the quintessence value,
the angular power spectrum looks very distorted compared to
the structure required by the data.
It is noteworthy that all curves are computed for the same background model,
so that the same relations between the redshift and luminosity distance or
the angular diameter distance are obtained.

\begin{figure}
\begin{center}
\begin{minipage}{12cm}
\vspace*{-25pt}
\hspace*{-20pt}\includegraphics[width=12.0cm]{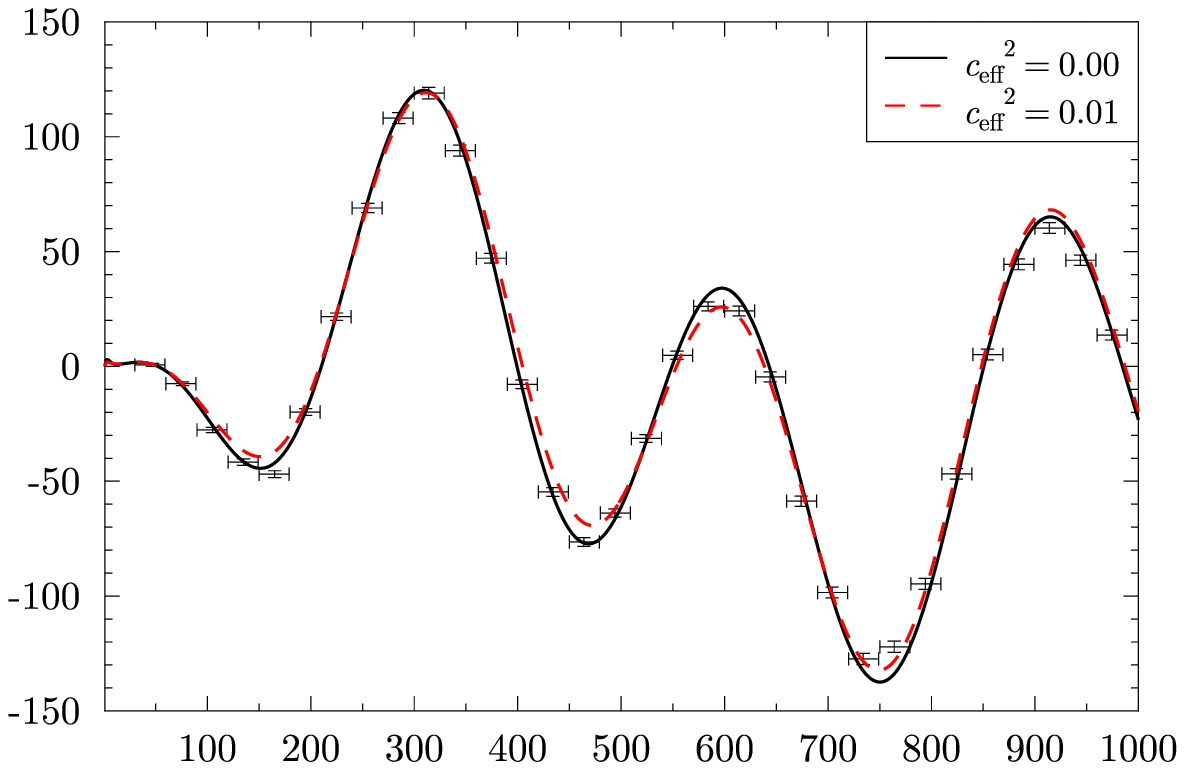}
\put(-34,22){$l$}
\put(-350,175){${\cal D}_l^{\hbox{\scriptsize TE}}$}
\put(-348,150){[$\mu\hbox{K}^2$]}
\put(-270,170){(a)}
\end{minipage}
\begin{minipage}{12cm}
\vspace*{-45pt}
\hspace*{-20pt}\includegraphics[width=12.0cm]{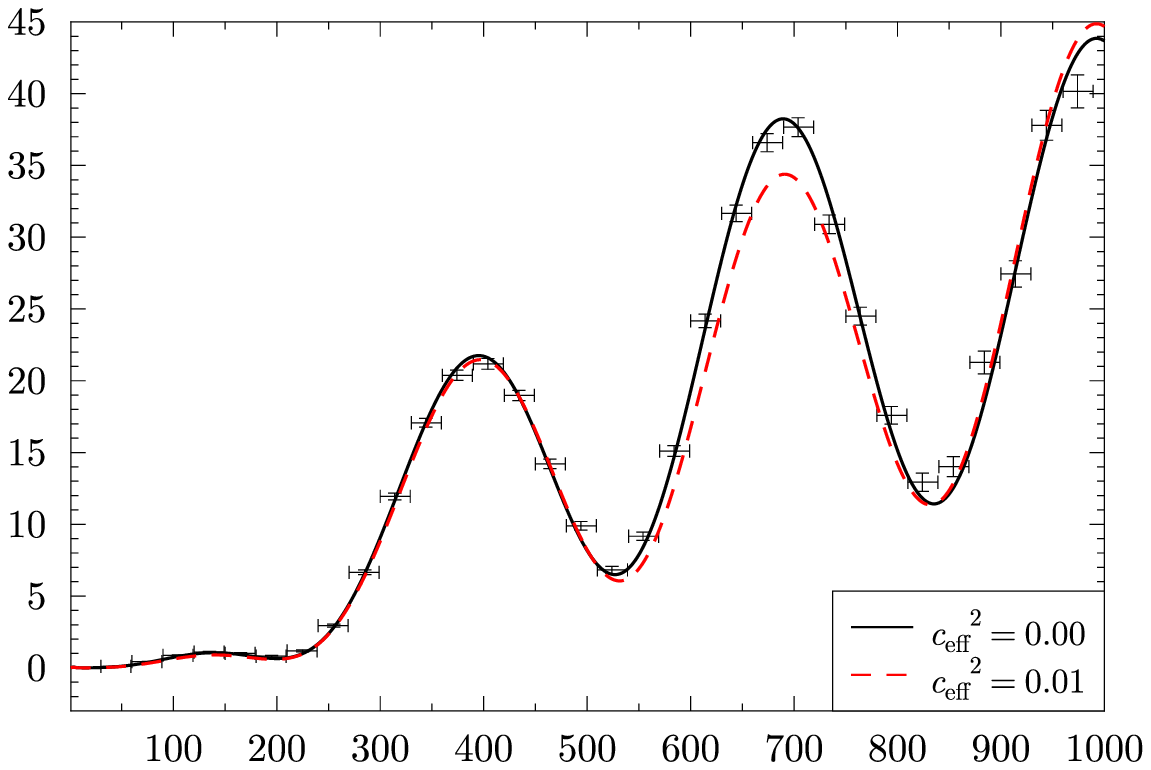}
\put(-34,22){$l$}
\put(-350,175){${\cal D}_l^{\hbox{\scriptsize EE}}$}
\put(-348,150){[$\mu\hbox{K}^2$]}
\put(-270,170){(b)}
\end{minipage}
\vspace*{-25pt}
\end{center}
\caption{\label{Fig:Tl_best_fit_planck_binned_pol}
The polarization power spectra
${\cal D}_l^{\hbox{\scriptsize TE}} = l(l+1) C^{\hbox{\scriptsize TE}}_l/(2\pi)$
and
${\cal D}_l^{\hbox{\scriptsize EE}} = l(l+1) C^{\hbox{\scriptsize EE}}_l/(2\pi)$
are shown in panels (a) and (b), respectively,
for the best-fit model as stated in the text.
The binned TE and EE data of the Planck 2015 data release
\cite{Planck_2015_I} are plotted with the corresponding error bars.
In addition, the polarization power spectra are shown for the model
with the slightly increased effective speed of sound
$c^2_{\hbox{\scriptsize eff}}=0.01$.
}
\end{figure}

Our MCMC algorithm computes the probability
needed for the model sequence only from
the temperature angular power spectrum ${\cal D}_l^{\hbox{\scriptsize TT}}$,
so the match with the polarization data is completely ignored
for the generation of the MCMC sequence.
However, it is nevertheless important to compare the Planck
polarization data \cite{Planck_2015_I} with the spectrum
of the best-fit model.
In figure \ref{Fig:Tl_best_fit_planck_binned_pol},
the polarization power spectra
${\cal D}_l^{\hbox{\scriptsize TE}} := l(l+1) C^{\hbox{\scriptsize TE}}_l/(2\pi)$
and
${\cal D}_l^{\hbox{\scriptsize EE}} := l(l+1) C^{\hbox{\scriptsize EE}}_l/(2\pi)$
are shown in panels (a) and (b), respectively,
in comparison to the binned TE and EE data of the Planck 2015 data release
\cite{Planck_2015_I}.
The best-fit model with $c^2_{\hbox{\scriptsize eff}}=0$ is plotted
as a full curve,
while the model with the slightly enhanced value of
$c^2_{\hbox{\scriptsize eff}}=0.01$ is shown as the dashed curve.
With respect to the fact that the best-fit model is solely determined
from the temperature power spectrum ${\cal D}_l^{\hbox{\scriptsize TT}}$,
it is assuring that also the polarization data agree well with
the best-fit model.
The comparison with the model with $c^2_{\hbox{\scriptsize eff}}=0.01$
reveals discrepancies as it is the case for the
temperature power spectrum ${\cal D}_l^{\hbox{\scriptsize TT}}$ shown in
figure \ref{Fig:Tl_best_fit_planck_binned}.
The largest deviations can be found for the EE spectrum around
$l\simeq 700$.

\begin{figure}
\begin{center}
\begin{minipage}{12cm}
\vspace*{-25pt}
\hspace*{-20pt}\includegraphics[width=12.0cm]{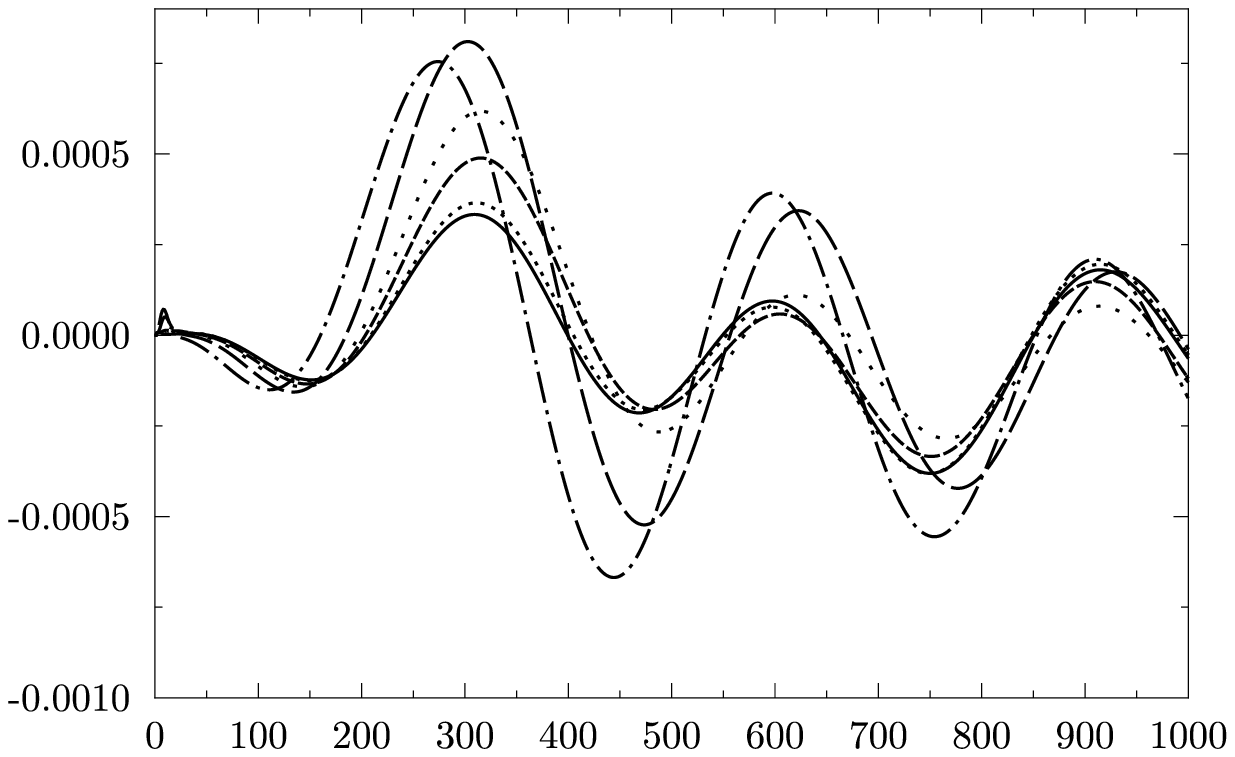}
\put(-34,22){$l$}
\put(-360,175){${\cal D}_l^{\hbox{\scriptsize TE}}$}
\put(-358,150){[$\mu\hbox{K}^2$]}
\put(-270,170){(a)}
\hspace*{-286pt}\begin{minipage}{3.5cm}
\vspace*{-132pt}\includegraphics[width=3.5cm]{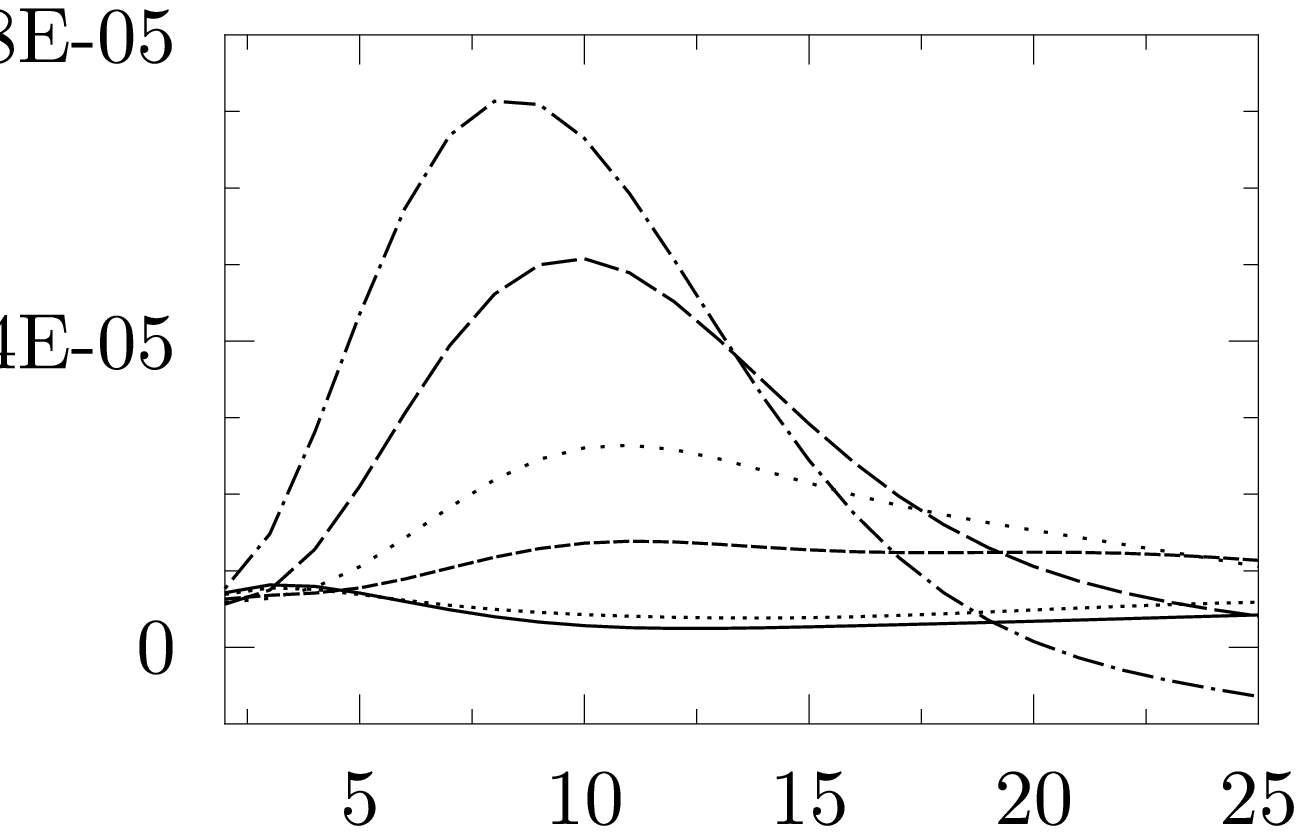}
\end{minipage}
\end{minipage}
\begin{minipage}{12cm}
\vspace*{-45pt}
\hspace*{-20pt}\includegraphics[width=12.0cm]{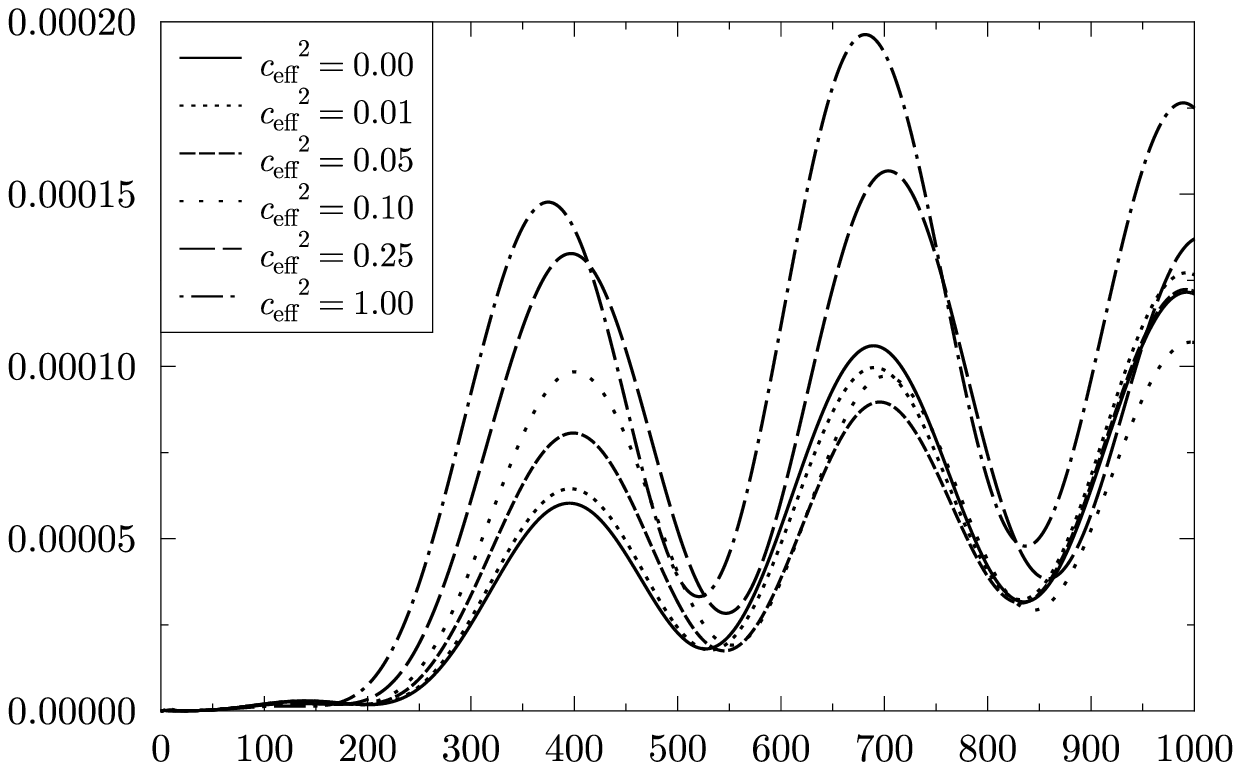}
\put(-34,22){$l$}
\put(-360,175){${\cal D}_l^{\hbox{\scriptsize EE}}$}
\put(-358,150){[$\mu\hbox{K}^2$]}
\put(-220,170){(b)}
\end{minipage}
\vspace*{-25pt}
\end{center}
\caption{\label{Fig:Tl_best_fit_pol_c2eff}
The polarization power spectra
${\cal D}_l^{\hbox{\scriptsize TE}}$ and ${\cal D}_l^{\hbox{\scriptsize EE}}$
are shown in panels (a) and (b), respectively,
in dependence on the effective speed of sound $c^2_{\hbox{\scriptsize eff}}$.
With the exception of $c^2_{\hbox{\scriptsize eff}}$,
all other cosmological parameters
are held fixed as those of the best-fit model.
Since no fit to the Planck data is carried out, these two panels reveal the
effect of $c^2_{\hbox{\scriptsize eff}}$ on the polarization power spectra.
The legend of panel (b) applies also to panel (a).
}
\end{figure}

As discussed in the case of the temperature power spectrum,
the fit to the Planck data hides the genuine effect of the
effective speed of sound $c^2_{\hbox{\scriptsize eff}}$ on the
polarization power spectra.
Thus, in contrast to figure \ref{Fig:Tl_best_fit_planck_binned_pol},
the figure \ref{Fig:Tl_best_fit_pol_c2eff} shows the
polarization power spectra ${\cal D}_l^{\hbox{\scriptsize TE}}$ and
${\cal D}_l^{\hbox{\scriptsize EE}}$ for the same set of cosmological
parameters with the exception of $c^2_{\hbox{\scriptsize eff}}$
which is varied from zero to one.
As discussed above,
the temperature power spectrum shown in figures \ref{Fig:Tl_as_fct_of_c_eff}
and \ref{Fig:Tl_as_fct_of_c_eff_log} is most sensitive
at and below the first acoustic peak to small changes
in $c^2_{\hbox{\scriptsize eff}}$ for $c^2_{\hbox{\scriptsize eff}}\lesssim 0.1$.
This behaviour differs from the polarization spectra.
The inspection of figure \ref{Fig:Tl_best_fit_pol_c2eff} reveals
the general trend that the polarization amplitude increases
with increasing $c^2_{\hbox{\scriptsize eff}}$.
This increase in polarization power is also noticeable around the
second peak of the EE spectrum.
However, the systematic increase is not observed around the third peak.
The cross-power spectrum ${\cal D}_l^{\hbox{\scriptsize TE}}$
displays a signature of the value of $c^2_{\hbox{\scriptsize eff}}$
for $l\lesssim 20$ as shown by the inset of figure
\ref{Fig:Tl_best_fit_pol_c2eff}(a)
which appears in the main figure as a small spike.

\section{Summary}

In this paper, the CMB anisotropy is computed for a phenomenological
unified dark matter (UDM) model, where the behaviour of the cold dark matter
and of the cosmological constant of the $\Lambda$CDM model
is represented by a single component.
At early times the UDM behaviour mimics that of cold dark matter,
and thus, this component might be called early-matter-like dark energy.
It changes the cosmological background model by modifying
the scale factor already at early times.
This in turn influences the growth of the perturbations
of the metric and the various energy components.
In order to compute the CMB anisotropy, an expansion of the scale factor
in terms of the conformal time is derived.
It is used for a power series expansion in terms of the conformal time
of the metric perturbations as well as
of the energy perturbations for all components.
These series expansions are given in the Appendix.
This allows to start the numerical integration of the Boltzmann hierarchy
for the computation of the CMB anisotropy at sufficiently late times,
since the start values for the numerical integration are computed
from the corresponding power series.
In this way the isentropic initial conditions are accurately taken into account.
Simultaneously, this procedure circumvents numerical accuracy problems
at very small conformal times.

The algorithm is applied to the phenomenological arctan-UDM scenario
which is analysed in \cite{Cuzinatto_Medeiros_deMorais_2014}.
This model is defined by the equation of state (\ref{Def:model_eos})
which is parameterised by two coefficients $\alpha$ and $\beta$.
The UDM model is compared in \cite{Cuzinatto_Medeiros_deMorais_2014}
with supernovae Ia data, the baryon acoustic oscillations and gamma-ray data,
and it is found that a good agreement is achieved for $\alpha\simeq 2.14$
and $\beta\simeq0.95$.
Since their analysis does not discuss the CMB anisotropy,
we apply our CMB algorithm to their UDM model in this work.
It is indeed found that there is a parameter range for which
the CMB anisotropy agrees with the Planck 2015 data \cite{Planck_2015_I},
if one allows for a vanishing effective speed of sound
$c^2_{\hbox{\scriptsize eff}}$
which is defined in (\ref{Eq:gdm_etropy_perturbation}).
This is consistent with the lesson from the generalised Chaplygin gas model,
which predicts a wrong evolution of the perturbations except one allows
the generation of entropy perturbations.

In \cite{Planck_2015_XIV} a great plethora of dark energy models is analysed
with respect to the Planck 2015 data.
One important implication is that the density of dark energy at early times
has to be below 2\% of the critical density using only the Planck data.
Taking other data into account, \cite{Planck_2015_XIV} reports
the even lower bound $\Omega_{\hbox{\scriptsize de}}(\eta)<0.0071$
(95\% CL, Planck TT+lensing+BAO+SNe+$H_0$) at early times.
Although the arctan-UDM model has a significantly larger value of
$\Omega_{\hbox{\scriptsize de}}(\eta)$ at early times,
it elegantly escapes this bound, because its equation of state
(\ref{Def:model_eos}) has already changed to that of a CDM component
at the relevant times.
This loophole is emphasised by the
figures \ref{Fig:Tl_best_fit_planck} and \ref{Fig:Tl_best_fit_planck_binned},
which reveal the good match between the theoretical prediction and the
Planck 2015 observations of the temperature CMB angular power spectrum.
For the same set of cosmological parameters, the arctan-UDM model also
predicts polarization spectra that agree with the Planck data as shown
in figure \ref{Fig:Tl_best_fit_planck_binned_pol}.
This agreement is, however, spoiled if the effective speed of sound
$c^2_{\hbox{\scriptsize eff}}$ is not very close to zero.
So the conclusion is that UDM models can match the observed CMB anisotropy
if the effective speed of sound is almost zero.

\vspace*{10pt}
{\bf \large Appendix}
\appendix

\section{The power series of the potentials in conformal time}
\label{app:potentials}

The evolution equations of the Newtonian metric perturbations $\Psi$
and $\Phi$ are given in Fourier space by (see, e.\,g.\ \cite{Hu_1995})
\begin{equation}
\label{Eq:Metric_TT}
3 \left(\frac{a'(\eta)}{a(\eta)}\right)^2\,\Psi(\eta) \, - \,
3 \frac{a'(\eta)}{a(\eta)}\,\Phi'(\eta) \, - \, K_2 k^2 \Phi
\; = \;
-\frac{4\pi G}{c^2} \, a^2(\eta) \rho_T(\eta) \delta_T(\eta)
\hspace{10pt} ,
\end{equation}
\begin{equation}
\label{Eq:Metric_ST}
\frac{a'(\eta)}{a(\eta)}\,\Psi(\eta) - \Phi'(\eta)  \; = \;
\frac{4\pi G}{c^2} \, a^2(\eta) \Big(\rho_T(\eta)+p_T(\eta)/c^2\Big)
\frac{v_T(\eta)}k
\hspace{10pt} ,
\end{equation}
and
\begin{equation}
\label{Eq:Metric_SS}
k^2 (\Psi(\eta)+\Phi(\eta)) \; = \;
-\frac{8\pi G}{c^4} \, a^2(\eta) p_T(\eta) \Pi_T(\eta)
\hspace{10pt} .
\end{equation}
Here, $\rho_T(\eta)$, $p_T(\eta)$, $v_T(\eta)$, and $\Pi_T(\eta)$ are
the matter density, the pressure, the velocity, and the anisotropic stress
perturbation in the total matter gauge, respectively,
as defined in \cite{Hu_1995}.
Furthermore, the quantity $K_2$ in eq.\,(\ref{Eq:Metric_TT}) is
defined by
$$
K_l \; := \; 1-(l^2-1)K/k^2
\hspace{10pt} \hbox{ with the curvature constant} \hspace{10pt}
K \in \{-1,0,+1\}
\hspace{10pt} ,
$$
for a hyperbolic, flat, or spherical space, respectively.
The wave number $k$ is related to the eigenvalue $E$ of the Laplace-Beltrami
operator by $k=\sqrt{E}=\sqrt{\beta^2-K}$.
In the case of positive curvature, the eigenvalue spectrum is discrete 
$\beta=3,4,5,\dots$.
For a flat space $K=0$ or for a negatively curved space $K=-1$,
one has $\beta\in {\mathbb R}^+$.

The potentials are expanded up to second order
\begin{equation}
\label{Eq:phi_expansion}
\Phi(\eta) \; = \;
\Phi_0 \, + \, \Phi_1 \eta \, + \, \frac 12 \Phi_2 \eta^2 \, + \, O(\eta^3)
\end{equation}
and
\begin{equation}
\label{Eq:psi_expansion}
\Psi(\eta) \; = \;
\Psi_0 \, + \, \Psi_1 \eta \, + \, \frac 12 \Psi_2 \eta^2 \, + \, O(\eta^3)
\hspace{10pt} .
\end{equation}
Imposing the isentropic initial conditions (\ref{Eq:initial_condition}),
these expansion coefficients are obtained as
\begin{equation}
\label{Eq:psi_0}
\Psi_0 \; = \; - \frac{\Phi_0}{1+\frac 25f_\nu}
\hspace{10pt} \hbox{ with the neutrino energy fraction} \hspace{10pt}
f_\nu := \frac{\Omega_\nu}{\Omega_{\hbox{\scriptsize rad}}}
\end{equation}
and $\Phi_0$ is determined by the initial power spectrum.
The next order is derived as
\begin{equation}
\label{Eq:phi_1}
\Phi_1 \; = \;
\frac{\frac{a_2}{a_1} \left(1+\frac{16}{15} f_\nu\right) - \frac{16}{15} \frac{a_1^2}d (1-f_\nu)}
{4+\frac{8}{15} f_\nu} \; \Psi_0
\end{equation}
and
\begin{equation}
\label{Eq:psi_1}
\Psi_1 \; = \; 3 \Phi_1 \, - \, \frac{a_2}{a_1} \, \Psi_0
\hspace{10pt} .
\end{equation}
The potentials depend on the perturbations in the various energy components,
and thus, the derivation of (\ref{Eq:phi_1}) and (\ref{Eq:psi_1}) already
requires the solutions of the corresponding energy perturbation equations
in terms of the conformal time $\eta$.
In these equations occur the optical depth for which the approximation
$\dot\tau=d/a^2(\eta)$ is used where $d$ is a constant for
$\eta\ll\eta_{\hbox{\scriptsize rec}}$.

The second-order terms are more involved and we define the quantities
$$
Q_2 \; := \;
\frac 13 k^2 K_2 (2\Phi_0-\Psi_0) + 6 \frac{a_2}{a_1}(2\Phi_1-\Psi_1) +
\left(\frac 23k^2 - 6\frac{a_3}{a_1}\right)\Psi_0
$$
and
\begin{eqnarray} \nonumber
R_2 & := & -2 \left( \left(\frac{a_2}{a_1}\right)^2 + 2\frac{a_3}{a_1} \right)
(\Phi_0+\Psi_0) - 4 \frac{a_2}{a_1}(\Phi_1+\Psi_1)
\\ & & \nonumber \hspace{30pt} - \;
\frac{16}{15}(1-f_\nu) \frac{a_1}{d^2} \left\{ (2d a_2-11a_1^3)\Psi_0 +
4 d a_1 \Phi_1 \right\}
\\ & & \nonumber \hspace{30pt} + \;
\frac{k^2}{45} f_\nu \left( 1 + \frac 45 K_2 + \frac{27}{35} K_3\right)\Psi_0
\end{eqnarray}
in order to write
\begin{equation}
\label{Eq:phi_psi_2}
\Phi_2 \; = \; \frac{R_2 - (1+\frac 2{15}f_\nu) Q_2}{5+\frac 25 f_\nu}
\hspace{10pt} \hbox{ and } \hspace{10pt}
\Psi_2 \; = \; 4 \Phi_2 \, + \, Q_2
\hspace{10pt} .
\end{equation}
If the initial condition $\Phi_0$ is chosen,
the expansions (\ref{Eq:phi_expansion}) and (\ref {Eq:psi_expansion})
can thus be computed solely from background model data
that is from the coefficients $a_1$, $a_2$ and $a_3$.

\section{The generalised matter perturbations in terms of the potentials}
\label{app:matter_perturbation}

The differential equations for the generalised dark matter perturbations,
given in eqs.\,(\ref{Eq:gdm_density}) and (\ref{Eq:gdm_velocity}),
can be rewritten for the density perturbation as (see \cite{Hu_1998})
\begin{equation}
\label{Eq:Dgl_d_G}
\dot\delta_{\hbox{\scriptsize g}} \; = \;
-\, (1+w_{\hbox{\scriptsize g}})(kv_{\hbox{\scriptsize g}}+3\dot\Phi)
\, - \,
3 \frac{\dot a}a (c_{\hbox{\scriptsize eff}}^2-w_{\hbox{\scriptsize g}}) \,
\delta_{\hbox{\scriptsize g}}^{\hbox{\scriptsize (rest)}}
\, - \,
3\frac{\dot a}a \dot w_{\hbox{\scriptsize g}} \frac{v_{\hbox{\scriptsize g}}}k
\end{equation}
and for the velocity as
\begin{equation}
\label{Eq:Dgl_v_G}
\dot v_{\hbox{\scriptsize g}} \; = \;
-\, \frac{\dot a}a \,v_{\hbox{\scriptsize g}}
\, + \,
\frac{c_{\hbox{\scriptsize eff}}^2}{1+w_{\hbox{\scriptsize g}}}
k \delta_{\hbox{\scriptsize g}}^{\hbox{\scriptsize (rest)}}
\, + \, k \Psi
\hspace{10pt} ,
\end{equation}
where the density perturbation in the rest frame of the
generalised dark matter is defined in (\ref{Def:d_rest}).
The generalised dark matter perturbation is expanded as
\begin{equation}
\delta_{\hbox{\scriptsize g}}(\eta) \; = \;
\delta_{\hbox{\scriptsize g},0} \, + \, \delta_{\hbox{\scriptsize g},1} \eta
\, + \, \frac 12\delta_{\hbox{\scriptsize g},2} \eta^2 \, + \, O(\eta^3)
\end{equation}
and the coefficients are calculated as
\begin{equation}
\label{Eq:d_G_0_1}
\delta_{\hbox{\scriptsize g},0} \; = \;
-\, \frac 32 \, \Psi_0
\hspace{20pt} , \hspace{20pt}
\delta_{\hbox{\scriptsize g},1} \; = \;
-\, 3 \Phi_1 \, - \, \frac 32 \frac{a_1}{A_0} \, w_1 \, \Psi_0
\end{equation}
and
\begin{eqnarray}
\label{Eq:d_G_2}
\delta_{\hbox{\scriptsize g},2} & = &
-\,3 \left[ \Phi_2 + \frac{k^2}6\Psi_0 +
 \frac{c_{\hbox{\scriptsize eff}}^2}{4+3c_{\hbox{\scriptsize eff}}^2}
\,\frac{k^2}2\, K_2(2\Phi_0-\Psi_0) \right]
\\ &  & \nonumber
\, - \,
3 w_1 \left(\frac{a_2}{A_0}\Psi_0 + 2\frac{a_1}{A_0}\Phi_1 \right)
\, - \,
3 w_2 \frac{a_1^2}{A_0^2}\Psi_0
\hspace{10pt} .
\end{eqnarray}
These coefficients are derived for an equation of state as given
in (\ref{Def:Early-Matter-Like})
where also the coefficients $w_1$ and $w_2$ are defined.

Expanding the velocity perturbations analogously as
\begin{equation}
v_{\hbox{\scriptsize g}}(\eta) \; = \;
v_{\hbox{\scriptsize g},1} \eta
\, + \, \frac 12 v_{\hbox{\scriptsize g},2} \eta^2
\, + \, \frac 16 v_{\hbox{\scriptsize g},3} \eta^3 \, + \, O(\eta^4)
\end{equation}
leads with (\ref{Eq:Dgl_v_G}) to
\begin{equation}
\label{Eq:v_G_1_2}
v_{\hbox{\scriptsize g},1} \; = \; \frac k2 \, \Psi_0
\hspace{15pt} , \hspace{15pt}
v_{\hbox{\scriptsize g},2} \; = \; k \, \big(\Psi_1-\Phi_1\big)
\end{equation}
and
\begin{equation}
v_{\hbox{\scriptsize g},3} \; = \;
k \left[ \Psi_2-\Phi_2 - \frac{k^2}6 \left[\Psi_0 +
\frac{2-3 c_{\hbox{\scriptsize eff}}^2}{4+3c_{\hbox{\scriptsize eff}}^2}
K_2 (2\Phi_0-\Psi_0) \right] \right]
\hspace{10pt} .
\end{equation}
These expansions are valid for all non-interacting fluids and
those of the cold dark matter are obtained from the generalised dark matter
equations by setting
$c_{\hbox{\scriptsize eff}}^2=w_1=w_2=0$.
This leads to
\begin{equation}
\label{Eq:d_cdm_0_1_2}
\delta_{\hbox{\scriptsize cdm},0} \; = \; -\, \frac 32 \, \Psi_0
\hspace{6pt} , \hspace{6pt}
\delta_{\hbox{\scriptsize cdm},1} \; = \; -\, 3 \Phi_1
\hspace{6pt} \hbox{and} \hspace{6pt}
\delta_{\hbox{\scriptsize cdm},2} \; = \; -\, 3\Phi_2 - \frac{k^2}2 \Psi_0
\end{equation}
and
\begin{eqnarray}
\label{Eq:v_cdm_1_2_3}
v_{\hbox{\scriptsize cdm},1} & = & \frac k2 \, \Psi_0
\hspace{6pt} , \hspace{6pt}
v_{\hbox{\scriptsize cdm},2} \; = \; k \, \big(\Psi_1-\Phi_1\big)
\\ & & \nonumber
\hbox{ and } \hspace{10pt}
v_{\hbox{\scriptsize cdm},3} \; = \; 
k \left[ \Psi_2-\Phi_2 - \frac{k^2}6 \left[\Psi_0 +
\frac{K_2}2 (2\Phi_0-\Psi_0) \right] \right]
\hspace{10pt} .
\end{eqnarray}
The velocity equation of the baryonic matter differs from (\ref{Eq:Dgl_v_G})
by the additional interaction term
$\dot\tau(\Theta_1-v_{\hbox{\scriptsize bar}})/R$
with $R=3\varepsilon_{\hbox{\scriptsize bar}}/4\varepsilon_\gamma$
due to the Compton scattering.
Up to the considered order, the density perturbations are the same as
in the case of cold dark matter, eqs.\,(\ref{Eq:d_cdm_0_1_2}),
\begin{equation}
\label{Eq:d_bar_0_1_2}
\delta_{\hbox{\scriptsize bar},0} \; = \; -\, \frac 32 \, \Psi_0
\hspace{6pt} , \hspace{6pt}
\delta_{\hbox{\scriptsize bar},1} \; = \; -\, 3 \Phi_1
\hspace{6pt} \hbox{and} \hspace{6pt}
\delta_{\hbox{\scriptsize bar},2} \; = \; -\, 3\Phi_2 - \frac{k^2}2 \Psi_0
\hspace{10pt} ,
\end{equation}
but the third order term in the velocity perturbation differs
\begin{equation}
\label{Eq:v_bar_1_2_3}
v_{\hbox{\scriptsize bar},1} \; = \; \frac k2 \, \Psi_0
\hspace{6pt} , \hspace{6pt}
v_{\hbox{\scriptsize bar},2} \; = \; k \, \big(\Psi_1-\Phi_1\big)
\hspace{6pt} \hbox{and} \hspace{6pt}
v_{\hbox{\scriptsize bar},3} \; = \; \Theta_{1,3}
\end{equation}
with $\Theta_{1,3}$ given in (\ref{Eq:d_Theta1_3}).

\section{The radiation perturbations in terms of the potentials}
\label{app:radiation_perturbation}

From the Boltzmann hierarchy of the photon perturbations $\Theta_l$,
we need for our small $\eta$ expansion only the first four equations
(see \cite{Hu_1995})
\begin{eqnarray}
\label{Eq:Dgl_Theta0}
\dot\Theta_0 & = & -\, \frac k3 \Theta_1 \, - \, \dot\Phi
\hspace{10pt} ,
\\
\label{Eq:Dgl_Theta1}
\dot\Theta_1 & = &
k \left[ \Theta_0 + \Psi - \frac 25 \sqrt{K_2}\Theta_2\right] -
\dot\tau(\Theta_1-v_{\hbox{\scriptsize bar}})
\hspace{10pt} ,
\\
\label{Eq:Dgl_Theta2}
\dot\Theta_2 & = &
k \left[ \frac 23 \sqrt{K_2}\Theta_1 - \frac 37 \sqrt{K_3}\Theta_3\right] -
\dot\tau\left( \frac 9{10} \Theta_2 - \frac{\sqrt 6}{10} E_2 \right)
\hspace{10pt} ,
\\
\label{Eq:Dgl_Theta3}
\dot\Theta_3 & = &
k \left[ \frac 35 \sqrt{K_3}\Theta_2 - \frac 49 \sqrt{K_4}\Theta_4\right] -
\dot\tau \Theta_3
\hspace{10pt} .
\end{eqnarray}
The density perturbation of the photons is defined as
$\Theta_0:=\frac 14 \delta_\gamma$.
For the polarization, the equations of the first two $E$ modes are required
\begin{equation}
\label{Eq:Dgl_E2}
\dot E_2 \; = \; - k \frac{\sqrt 5}7 \sqrt{K_3} E_3 -
\dot\tau\left( \frac 4{10} E_2 - \frac{\sqrt 6}{10} \Theta_2 \right)
\end{equation}
and
\begin{equation}
\label{Eq:Dgl_E3}
\dot E_3 \; = \; k \left[\frac 1{\sqrt 5} \sqrt{K_3} E_2 -
\frac{\sqrt{12}}{9} \sqrt{K_4} E_4 \right] -
\dot\tau E_3
\hspace{10pt} .
\end{equation}
The perturbations $\Theta_l$ for $l=0,1,2,3$ are expanded as
\begin{equation}
\Theta_l(\eta) \; = \; \sum_{j=l}^{l+2} \frac{\Theta_{l,j}}{j!}\; \eta^j 
\, + \, O(\eta^{l+3})
\end{equation}
and of $E_l$ for $l=2,3$ as
\begin{equation}
E_l(\eta) \; = \; \sum_{j=l}^{l+2} \frac{E_{l,j}}{j!}\; \eta^j 
\, + \, O(\eta^{l+3})
\hspace{10pt} .
\end{equation}
The evaluation of the differential equations leads for
isentropic initial conditions for $l=0$ to
\begin{equation}
\label{Eq:d_Theta0_0_1_2}
\Theta_{0,0} \; = \; -\, \frac 12 \, \Psi_0
\hspace{9pt} , \hspace{9pt}
\Theta_{0,1} \; = \; -\, \Phi_1
\hspace{9pt} \hbox{and} \hspace{9pt}
\Theta_{0,2} \; = \; -\, \Phi_2 \, - \, \frac{k^2}6 \Psi_0
\end{equation}
for $l=1$ to
\begin{eqnarray}
\label{Eq:d_Theta1_1_2}
\Theta_{1,1} & = & \frac k2 \, \Psi_0
\hspace{9pt} , \hspace{9pt}
\Theta_{1,2} \; = \; k (\Psi_1-\Phi_1)
\\ & & \label{Eq:d_Theta1_3}
\hspace{80pt} \hbox{ and } \hspace{10pt}
\Theta_{1,3} \; = \; k \left[ \Psi_2-\Phi_2 - \frac{k^2}6 \Psi_0 \right]
\end{eqnarray}
and for $l=2$ to
\begin{eqnarray}
\label{Eq:d_Theta2_2_3_4}
\Theta_{2,2} & = & 0
\hspace{9pt} , \hspace{9pt}
\Theta_{2,3} \; = \; \frac 83 k^2 \sqrt{K_2}\, \frac{a_1^2}d \Psi_0
\hspace{9pt} \hbox{and} \hspace{9pt}
\\ \nonumber
\Theta_{2,4} & = &
\frac{16 k^2\sqrt{K_2} a_1}{3d^2} \left\{
(2 d a_2 - 11 a_1^3)\Psi_0 + 4 d a_1 \Phi_1 \right\}
\hspace{10pt} .
\end{eqnarray}
For the $E$ mode, one finds for the relevant non-vanishing coefficients
$$
E_{2,3} \; = \;
\sqrt{\frac 83}\, \frac{k^2\sqrt{K_2} a_1^2}{d}\, \Psi_0
\hspace{9pt} \hbox{and} \hspace{9pt}
E_{2,4} \; = \; 
\frac{8 \sqrt{\frac 23} k^2\sqrt{K_2} a_1}{d^2} \left\{
(d a_2 - 13 a_1^3)\Psi_0 + 2 d a_1 \Phi_1 \right\}
\hspace{10pt} .
$$

The expansions for the neutrino perturbations are defined analogously
to the photons
\begin{equation}
N_l(\eta) \; = \; \sum_{j=l}^{l+2} \frac{N_{l,j}}{j!}\; \eta^j 
\, + \, O(\eta^{l+3})
\hspace{10pt} .
\end{equation}
The differential equations for the neutrinos are obtained
from (\ref{Eq:Dgl_Theta0}) to (\ref{Eq:Dgl_Theta3}) by setting $\dot\tau=0$,
which lead for isentropic initial conditions for $l=0$ to
\begin{equation}
\label{Eq:d_N0_0_1_2}
N_{0,0} \; = \; -\, \frac 12 \, \Psi_0
\hspace{9pt} , \hspace{9pt}
N_{0,1} \; = \; -\, \Phi_1
\hspace{9pt} \hbox{and} \hspace{9pt}
N_{0,2} \; = \; -\, \Phi_2 \, - \, \frac{k^2}6 \Psi_0
\end{equation}
for $l=1$ to
\begin{eqnarray}
\label{Eq:d_N1_1_2_3}
N_{1,1} & = & \frac k2 \, \Psi_0
\hspace{10pt} , \hspace{10pt}
N_{1,2} \; = \; k (\Psi_1-\Phi_1)
\hspace{9pt} \hbox{and} \hspace{9pt}
\\ \nonumber
N_{1,3} & = &
k \left[ \Psi_2-\Phi_2 - \frac{k^2}6
\left(1 + \frac 45 K_2 \right)  \Psi_0 \right]
\end{eqnarray}
and for $l=2$ to
\begin{eqnarray}
\label{Eq:d_N2_2_3_4}
N_{2,2} & = & \frac 13 k^2 \sqrt{K_2} \Psi_0
\hspace{10pt} , \hspace{10pt}
N_{2,3} \; = \; \frac 23 k^2 \sqrt{K_2} (\Psi_1-\Phi_1)
\hspace{9pt} \hbox{and} \hspace{9pt}
\\ \nonumber
N_{2,4} & = &
\frac 23 k^2\sqrt{K_2} \left[ \Psi_2-\Phi_2 \, - \,
k^2 \left\{ \frac 16 + \frac 2{15}K_2 + \frac 9{70}K_3 \right\}
\Psi_0 \right]
\hspace{10pt} ,
\end{eqnarray}
and for $l=3$ to
\begin{equation}
N_{3,3} \; = \; \frac 15\,k^3 \sqrt{K_2}\sqrt{K_3} \, \Psi_0
\hspace{10pt} .
\end{equation}

\section{Isentropic initial condition}
\label{app:isentropic_initial_condition}

\begin{figure}
\begin{center}
\begin{minipage}{12cm}
\vspace*{-10pt}
\hspace*{-20pt}\includegraphics[width=12.0cm]{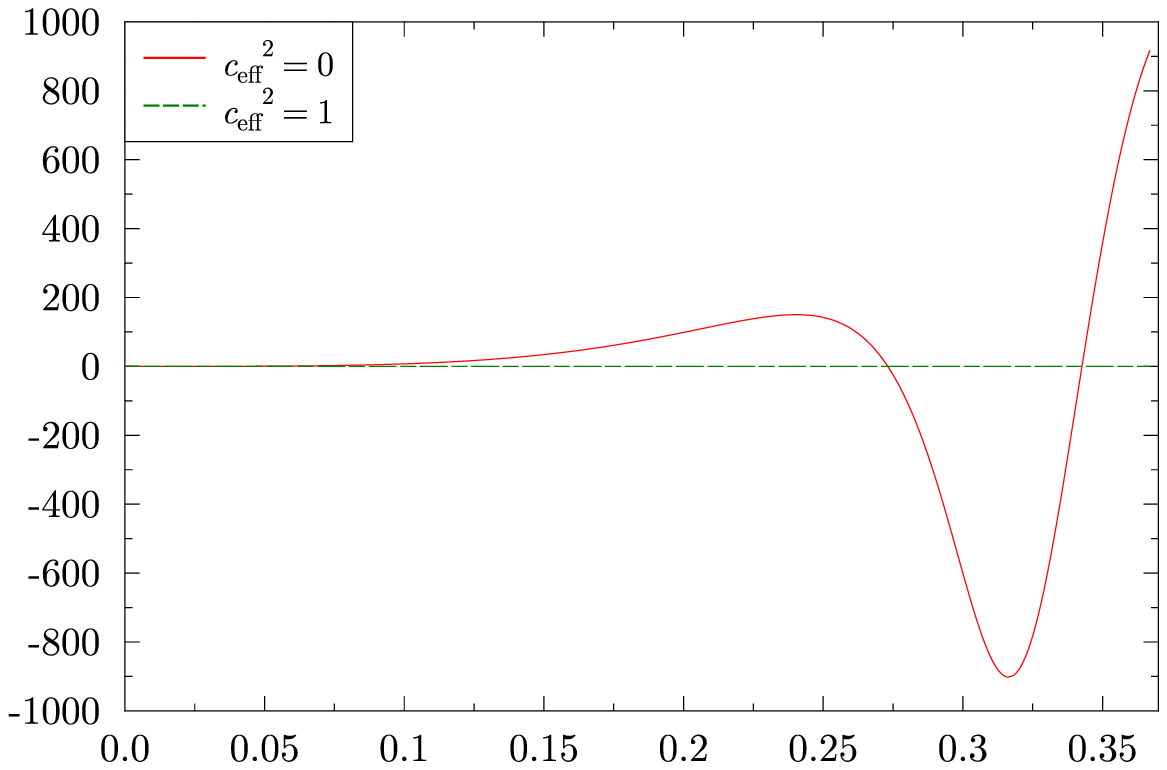}
\put(-40,22){$\eta$}
\put(-340,175){$w_{\hbox{\scriptsize g}}\Gamma$}
\end{minipage}
\vspace*{-25pt}
\end{center}
\caption{\label{Fig:w_Gamma}
The weighted entropy perturbation $w_{\hbox{\scriptsize g}}\Gamma$
of the UDM component is plotted for the two cases $c^2_{\hbox{\scriptsize eff}}=0$
and $c^2_{\hbox{\scriptsize eff}}=1$ for the wave number $k=2000$
using the best-fit model parameters.
}
\end{figure}

\begin{figure}
\begin{center}
\begin{minipage}{12cm}
\vspace*{-10pt}
\hspace*{-20pt}\includegraphics[width=12.0cm]{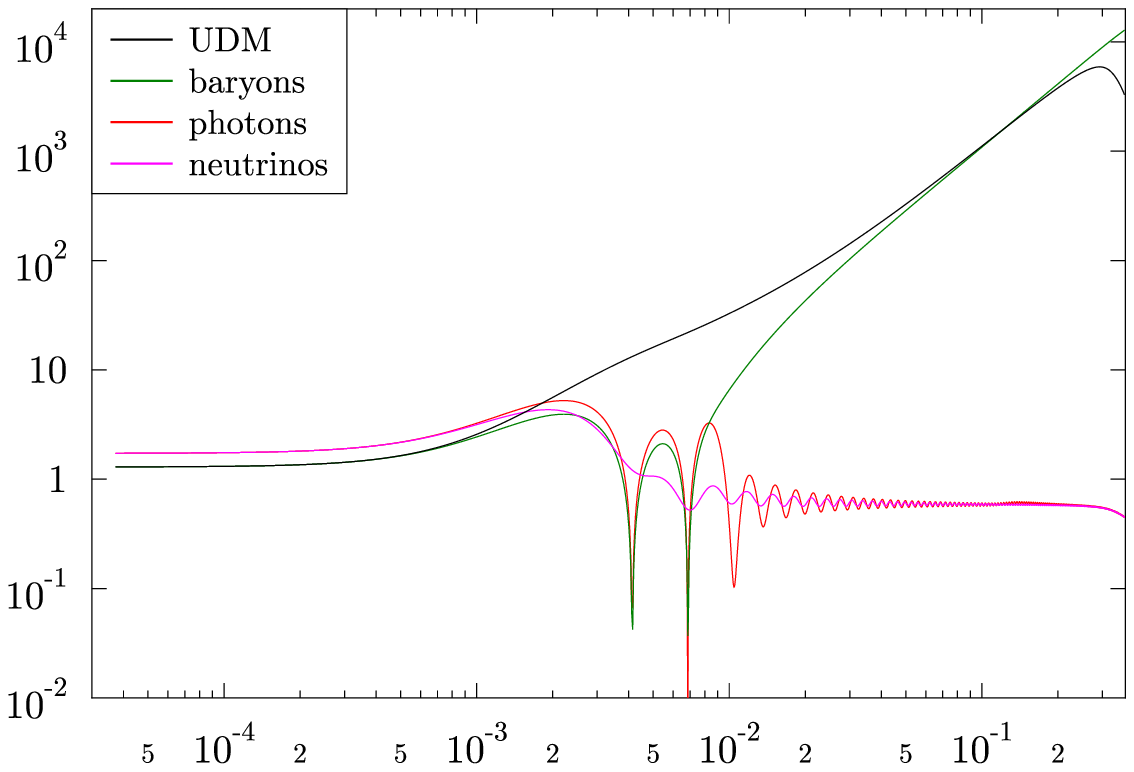}
\put(-40,22){$\eta$}
\put(-330,175){$|\delta_x|$}
\put(-210,160){(a)}
\put(-100,120){$c^2_{\hbox{\scriptsize eff}}=0$}
\end{minipage}
\begin{minipage}{12cm}
\vspace*{-30pt}
\hspace*{-20pt}\includegraphics[width=12.0cm]{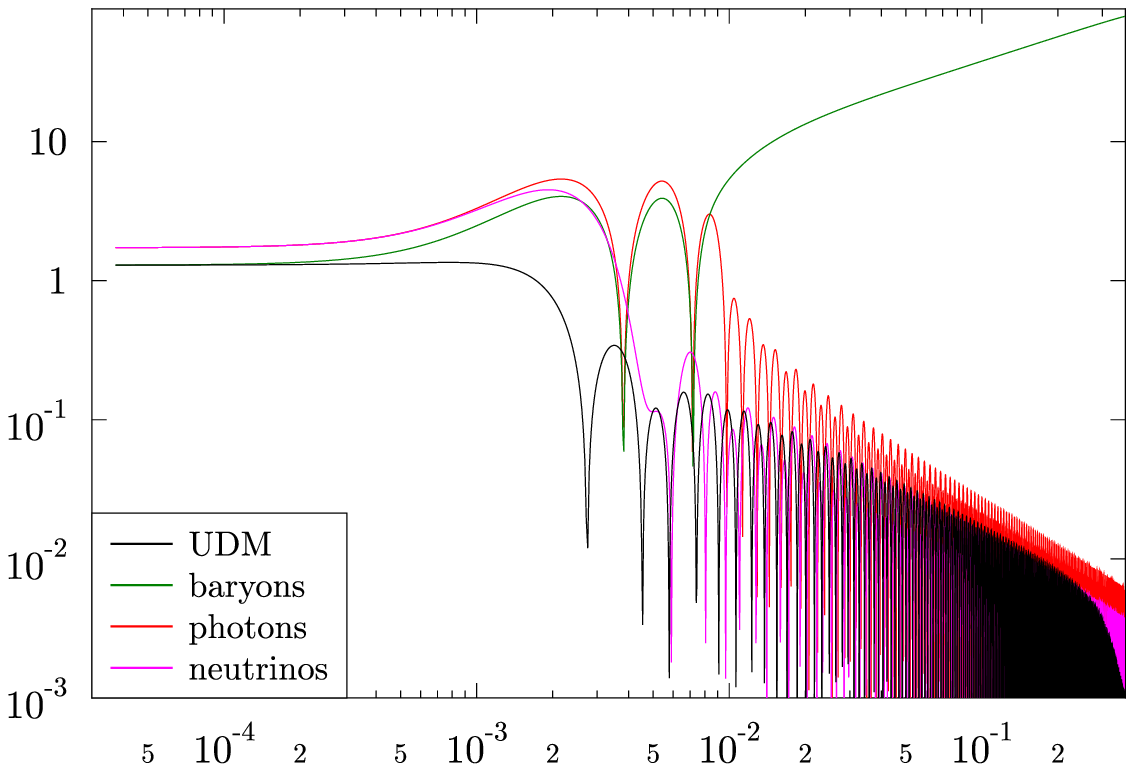}
\put(-40,22){$\eta$}
\put(-330,175){$|\delta_x|$}
\put(-260,160){(b)}
\put(-100,120){$c^2_{\hbox{\scriptsize eff}}=1$}
\end{minipage}
\vspace*{-25pt}
\end{center}
\caption{\label{Fig:perturbation_c2eff}
The modulus of the relative energy density perturbations
$|\delta_x|$ of the various components is plotted in dependence on
conformal time $\eta$.
The panels (a) and (b) show the results for $c^2_{\hbox{\scriptsize eff}}=0$
and $c^2_{\hbox{\scriptsize eff}}=1$, respectively.
The cosmological parameters are those of the best-fit model
leading to a recombination time $\eta_{\hbox{\scriptsize rec}}=0.00745$,
and the wave number $k=2000$ is used.
}
\end{figure}

In this appendix, it is demonstrated that the initial condition
(\ref{Eq:initial_condition}) implies a vanishing entropy perturbation
$\Gamma$ for $\eta\to 0$, so that the initial condition is a
genuine isentropic initial condition
without an isocurvature contribution.
From the definition (\ref{Eq:gdm_etropy_perturbation})
$$
w_{\hbox{\scriptsize g}} \Gamma \; = \;
\left(c^2_{\hbox{\scriptsize eff}}-c^2_{\hbox{\scriptsize g}}\,\right)
\,\delta_{\hbox{\scriptsize g}}^{\hbox{\scriptsize (rest)}}
$$
follows that one has to show
$\lim_{\eta\to 0}\delta_{\hbox{\scriptsize g}}^{\hbox{\scriptsize (rest)}}= 0$,
if $\lim_{\eta\to 0}\left(c^2_{\hbox{\scriptsize eff}}-
c^2_{\hbox{\scriptsize g}}\right)$ is finite and not necessarily equal to zero.
For the UDM model discussed in this paper,
one has $\lim_{\eta\to 0}\left(c^2_{\hbox{\scriptsize eff}}-
c^2_{\hbox{\scriptsize g}}\right) = c^2_{\hbox{\scriptsize eff}}$
because of $\lim_{\eta\to 0}c^2_{\hbox{\scriptsize g}}=w_{\hbox{\scriptsize g}}(0)=w_0=0$
and $0\leq c^2_{\hbox{\scriptsize eff}}\leq 1$.
To see the vanishing of $\delta_{\hbox{\scriptsize g}}^{\hbox{\scriptsize (rest)}}$,
one notes from the preceding appendices
$\delta_{\hbox{\scriptsize g}} = - \frac 32 (1+w_0)\Psi_0 + O(\eta)$,
$v_{\hbox{\scriptsize g}} = \frac k2 \Psi_0 \eta + O(\eta^2)$ and
$w_{\hbox{\scriptsize g}} = w_0 + O(\eta)$,
which determine the behaviour at early times of
$\delta_{\hbox{\scriptsize g}}^{\hbox{\scriptsize (rest)}}$
given in (\ref{Def:d_rest})
\begin{eqnarray}
\nonumber
\delta_{\hbox{\scriptsize g}}^{\hbox{\scriptsize (rest)}} & = & 
\delta_{\hbox{\scriptsize g}} \, + \,
3 \frac{\dot a}a\, (1+w_{\hbox{\scriptsize g}}) \, \frac{v_{\hbox{\scriptsize g}}}k
\\ & = & \nonumber
- \frac 32 (1+w_0)\Psi_0 \, + \,
3\, \frac 1\eta\, (1+w_0)\, \frac 1k \, \frac k2 \Psi_0 \eta
\, + \, O(\eta)
\; = \; O(\eta)
\hspace{10pt} .
\end{eqnarray}
This shows that no entropy perturbation is initially present
also for $w_0\neq 0$.
This is confirmed in figure \ref{Fig:w_Gamma} which shows the 
weighted entropy perturbation $w_{\hbox{\scriptsize g}}\Gamma$ for the
best-fit model using the wave number $k=2000$.
For small values of $\eta$ the quantity $w_{\hbox{\scriptsize g}}\Gamma$
approaches zero.
The significant difference between the two extreme cases
$c^2_{\hbox{\scriptsize eff}}=0$ and $c^2_{\hbox{\scriptsize eff}}=1$
occurs at later times and is due to the fact
that the energy perturbation of the unified dark matter component
increases in the first case but decays in the second case.
The different behaviour of the unified dark matter is visualised
together with the various other components in
figure \ref{Fig:perturbation_c2eff},
where the modulus of the relative energy density perturbations $|\delta_x|$
is plotted.
Note that figure \ref{Fig:perturbation_c2eff} uses a logarithmic scaling
in contrast to the linear scaling in figure \ref{Fig:w_Gamma}.
The energy density perturbations of the baryonic and the UDM component
are almost identical at small values of the conformal time $\eta$.
The initial perturbations of the photonic and neutrino components are larger
than those of the matter components due to the
condition (\ref{Eq:initial_condition}).
For the case $c^2_{\hbox{\scriptsize eff}}=0$,
shown in figure \ref{Fig:perturbation_c2eff}(a),
the UDM energy perturbation increases and the baryonic density perturbation
approaches after the recombination that of the UDM component.
This leads to the large baryonic clustering necessary for the structure
formation as in the usual $\Lambda$CDM model.
The case with $c^2_{\hbox{\scriptsize eff}}=1$,
shown in figure \ref{Fig:perturbation_c2eff}(b),
possesses a decaying UDM energy perturbation and the baryonic energy
perturbation increases only with a much lower rate after the recombination
as the comparison with panel (a) reveals.
This leads to the different behaviour of the entropy at late times
as seen in figure \ref{Fig:w_Gamma}.

\acknowledgments

The Planck 2015 data \cite{Planck_2015_I} from the LAMBDA website
(lambda.gsfc.nasa.gov) were used in this work.


\bibliographystyle{JHEP}
\bibliography{../bib_astro}

\providecommand{\href}[2]{#2}\begingroup\raggedright\begin{thebibliography}{10}

\bibitem{Hu_1998}
W.~{Hu}, {\it Structure formation with generalized dark matter},  {\em \apj}
  {\bf 506} (1998) 485--494,
  [\href{http://arxiv.org/abs/arXiv:astro-ph/9801234}{{\tt
  {arXiv:astro-ph/9801234}}}].

\bibitem{Kamenshchik_Moschella_Pasquier_2001}
A.~{Kamenshchik}, U.~{Moschella}, and V.~{Pasquier}, {\it {An alternative to
  quintessence}},  {\em Physics Letters B} {\bf 511} (2001) 265--268,
  [\href{http://arxiv.org/abs/gr-qc/0103004}{{\tt gr-qc/0103004}}].

\bibitem{Fabris_Goncalves_deSouza__2001}
J.~C. {Fabris}, S.~V.~B. {Goncalves}, and P.~E. {de Souza}, {\it {Density
  perturbations in an Universe dominated by the Chaplygin gas}},  {\em Gen.\
  Rel.\ Grav.} {\bf 34} (2002) 53--63,
  [\href{http://arxiv.org/abs/gr-qc/0103083}{{\tt gr-qc/0103083}}].

\bibitem{Bilic_Tupper_Viollier_2002}
N.~{Bili{\'c}}, G.~B. {Tupper}, and R.~D. {Viollier}, {\it {Unification of dark
  matter and dark energy: the inhomogeneous Chaplygin gas}},  {\em Physics
  Letters B} {\bf 535} (2002) 17--21,
  [\href{http://arxiv.org/abs/astro-ph/0111325}{{\tt astro-ph/0111325}}].

\bibitem{Bento_Bertolami_Sen_2002}
M.~C. {Bento}, O.~{Bertolami}, and A.~A. {Sen}, {\it {Generalized Chaplygin
  gas, accelerated expansion, and dark-energy-matter unification}},  {\em \prd}
  {\bf 66} (2002) 043507, [\href{http://arxiv.org/abs/gr-qc/0202064}{{\tt
  gr-qc/0202064}}].

\bibitem{Carturan_Finelli_2003}
D.~{Carturan} and F.~{Finelli}, {\it {Cosmological effects of a class of fluid
  dark energy models}},  {\em \prd} {\bf 68} (2003) 103501,
  [\href{http://arxiv.org/abs/astro-ph/0211626}{{\tt astro-ph/0211626}}].

\bibitem{Makler_deOliveira_Waga_2003}
M.~{Makler}, S.~Q. {de Oliveira}, and I.~{Waga}, {\it {Constraints on the
  generalized Chaplygin gas from supernovae observations}},  {\em Physics
  Letters B} {\bf 555} (2003) 1--6,
  [\href{http://arxiv.org/abs/astro-ph/0209486}{{\tt astro-ph/0209486}}].

\bibitem{Sandvik_Tegmark_Zaldarriaga_2004}
H.~B. {Sandvik}, M.~{Tegmark}, M.~{Zaldarriaga}, and I.~{Waga}, {\it {The end
  of unified dark matter?}},  {\em \prd} {\bf 69} (2004) 123524,
  [\href{http://arxiv.org/abs/astro-ph/0212114}{{\tt astro-ph/0212114}}].

\bibitem{Piattella_2010}
O.~F. {Piattella}, {\it {The extreme limit of the generalised Chaplygin gas}},
  {\em \jcap} {\bf 3} (2010) 12, [\href{http://arxiv.org/abs/arXiv:0906.4430}
  {{\tt {arXiv:0906.4430 [astro-ph.CO]}}}].

\bibitem{Reis_Waga_Calvao_Joras_2003}
R.~R. {Reis}, I.~{Waga}, M.~O. {Calv{\~a}o}, and S.~E. {Jor{\'a}s}, {\it
  {Entropy perturbations in quartessence Chaplygin models}},  {\em \prd} {\bf
  68} (2003) 061302, [\href{http://arxiv.org/abs/astro-ph/0306004}{{\tt
  astro-ph/0306004}}].

\bibitem{Scherrer_2004}
R.~J. {Scherrer}, {\it {Purely Kinetic k Essence as Unified Dark Matter}},
  {\em \prl} {\bf 93} (2004) 011301,
  [\href{http://arxiv.org/abs/astro-ph/0402316}{{\tt astro-ph/0402316}}].

\bibitem{Erickson_Caldwell_Steinhardt_Picon_Mukhanov_2002}
J.~K. {Erickson}, R.~R. {Caldwell}, P.~J. {Steinhardt}, C.~{Armendariz-Picon},
  and V.~{Mukhanov}, {\it {Measuring the Speed of Sound of Quintessence}},
  {\em \prl} {\bf 88} (2002) 121301,
  [\href{http://arxiv.org/abs/astro-ph/0112438}{{\tt astro-ph/0112438}}].

\bibitem{DeDeo_Caldwell_Steinhardt_2003}
S.~{DeDeo}, R.~R. {Caldwell}, and P.~J. {Steinhardt}, {\it {Effects of the
  sound speed of quintessence on the microwave background and large scale
  structure}},  {\em \prd} {\bf 67} (2003) 103509,
  [\href{http://arxiv.org/abs/astro-ph/0301284}{{\tt astro-ph/0301284}}].

\bibitem{Bertacca_Bartolo_Matarres_2010}
D.~{Bertacca}, N.~{Bartolo}, and S.~{Matarrese}, {\it {Unified Dark Matter
  Scalar Field Models}},  {\em Advances in Astronomy} {\bf 2010} (2010) 904379,
  [\href{http://arxiv.org/abs/arXiv:1008.0614}{{\tt
  {arXiv:1008.0614 [astro-ph.CO]}}}].

\bibitem{Hipolito_Velten_Zimdahl__2009}
W.~S. {Hip{\'o}lito-Ricaldi}, H.~E.~S. {Velten}, and W.~{Zimdahl}, {\it
  {Non-adiabatic dark fluid cosmology}},  {\em \jcap} {\bf 6} (2009) 16,
  [\href{http://arxiv.org/abs/arXiv:0902.4710}{{\tt
  {arXiv:0902.4710 [astro-ph.CO]}}}].

\bibitem{Piattella_Bertacc_Bruni_Pietrobon_2010}
O.~F. {Piattella}, D.~{Bertacca}, M.~{Bruni}, and D.~{Pietrobon}, {\it {Unified
  Dark Matter models with fast transition}},  {\em \jcap} {\bf 1} (2010) 14,
  [\href{http://arxiv.org/abs/arXiv:0911.2664}{{\tt
  {arXiv:0911.2664 [astro-ph.CO]}}}].

\bibitem{Bertacca_Bruni_Piattella_Pietrobon_2011}
D.~{Bertacca}, M.~{Bruni}, O.~F. {Piattella}, and D.~{Pietrobon}, {\it {Unified
  Dark Matter scalar field models with fast transition}},  {\em \jcap} {\bf 02}
  (2011) 018, [\href{http://arxiv.org/abs/arXiv:1011.6669}{{\tt
  {arXiv:1011.6669 [astro-ph.CO]}}}].

\bibitem{Campos_Fabris_Perez_Piattella_2013}
J.~P. {Campos}, J.~C. {Fabris}, R.~{Perez}, O.~F. {Piattella}, and H.~{Velten},
  {\it {Does Chaplygin gas have salvation?}},  {\em European Physical Journal
  C} {\bf 73} (2013) 2357, [\href{http://arxiv.org/abs/arXiv:1212.4136}
{{\tt {arXiv:1212.4136 [astro-ph.CO]}}}].

\bibitem{Borges_Carneiro_Fabris_Zimdahl_2013}
H.~A. {Borges}, S.~{Carneiro}, J.~C. {Fabris}, and W.~{Zimdahl}, {\it
  {Non-adiabatic Chaplygin gas}},  {\em Physics Letters B} {\bf 727} (2013)
  37--42, [\href{http://arxiv.org/abs/arXiv:1306.0917}{{\tt
  {arXiv:1306.0917 [astro-ph.CO]}}}].

\bibitem{Bruni_Lazkoz_Fernandez_2013}
M.~{Bruni}, R.~{Lazkoz}, and A.~{Rozas-Fern{\'a}ndez}, {\it {Phenomenological
  models for unified dark matter with fast transition}},  {\em \mnras} {\bf
  431} (2013) 2907--2916, [\href{http://arxiv.org/abs/arXiv:1210.1880}
  {{\tt {arXiv:1210.1880 [astro-ph.CO]}}}].

\bibitem{Carames_Fabris_Velten_2014}
T.~R.~P. {Caram{\^e}s}, J.~C. {Fabris}, and H.~E.~S. {Velten}, {\it {Spherical
  collapse for unified dark matter models}},  {\em \prd} {\bf 89} (2014)
  083533, [\href{http://arxiv.org/abs/arXiv:1401.5608}{{\tt
  {arXiv:1401.5608 [astro-ph.CO]}}}].

\bibitem{Kumar_Sen_2014}
S.~{Kumar} and A.~A. {Sen}, {\it {Clustering GCG: a viable option for unified
  dark matter-dark energy?}},  {\em \jcap} {\bf 10} (2014) 36,
  [\href{http://arxiv.org/abs/arXiv:1405.5688}{{\tt
  {arXiv:1405.5688 [astro-ph.CO]}}}].

\bibitem{Lazkoz_Leanizbarrutia_Salzano_2015}
R.~{Lazkoz}, I.~{Leanizbarrutia}, and V.~{Salzano}, {\it {Cosmological
  constrains on fast transition Unified Dark Matter models}},  {\em Journal of
  Physics Conference Series} {\bf 600} (2015) 012028.

\bibitem{Cuzinatto_Medeiros_deMorais_2014}
R.~R. {Cuzinatto}, L.~G. {Medeiros}, and E.~M. {de Morais}, {\it Observational
  constraints to a unified cosmological model},  {\em Astroparticle Physics}
  {\bf 73} (2016) 52--61, [\href{http://arxiv.org/abs/arXiv:1412.0145}
  {{\tt {arXiv:1412.0145 [astro-ph.CO]}}}].

\bibitem{Ma_Bertschinger_1995}
C.~{Ma} and E.~{Bertschinger}, {\it Cosmological perturbation theory in the
  synchronous and conformal {N}ewtonian gauges},  {\em \apj} {\bf 455} (1995)
  7--25.

\bibitem{Hu_1995}
W.~{Hu}, {\em {Wandering in the Background: A CMB Explorer}}.
\newblock PhD thesis, University of California at Berkeley, 1995.
\newblock \href{http://arxiv.org/abs/arXiv:astro-ph/9508126}{{\tt
  {arXiv:astro-ph/9508126}}}.

\bibitem{Mukhanov_Feldman_Brandenberger_1992}
V.~F. {Mukhanov}, H.~A. {Feldman}, and R.~H. {Brandenberger}, {\it Theory of
  cosmological perturbations},  {\em Physics Report} {\bf 215} (1992) 203--333.

\bibitem{Ballesteros_Lesgourgues_2010}
G.~{Ballesteros} and J.~{Lesgourgues}, {\it {Dark energy with non-adiabatic
  sound speed: initial conditions and detectability}},  {\em \jcap} {\bf 10}
  (2010) 14, [\href{http://arxiv.org/abs/arXiv:1004.5509}{{\tt
  {arXiv:1004.5509 [astro-ph.CO]}}}].

\bibitem{Planck_2015_I}
{Planck Collaboration}, R.~{Adam}, P.~A.~R. {Ade}, N.~{Aghanim}, Y.~{Akrami},
  M.~I.~R. {Alves}, M.~{Arnaud}, F.~{Arroja}, J.~{Aumont}, C.~{Baccigalupi},
  and et~al., {\it {Planck 2015 results. I. Overview of products and scientific
  results}},  \href{http://arxiv.org/abs/arXiv:1502.01582}{{\tt
  {arXiv:1502.01582 [astro-ph.CO]}}}.

\bibitem{Planck_Cosmo_Parameters_2013}
{Planck Collaboration}, P.~A.~R. {Ade}, N.~{Aghanim}, C.~{Armitage-Caplan},
  M.~{Arnaud}, M.~{Ashdown}, F.~{Atrio-Barandela}, J.~{Aumont},
  C.~{Baccigalupi}, A.~J. {Banday}, and et~al., {\it {Planck 2013 results. XVI.
  Cosmological parameters}},  {\em \aap} {\bf 571} (2014) A16,
  [\href{http://arxiv.org/abs/arXiv:1303.5076}{{\tt
  {arXiv:1303.5076 [astro-ph.CO]}}}].

\bibitem{Planck_2015_XIV}
{Planck Collaboration}, P.~A.~R. {Ade}, N.~{Aghanim}, M.~{Arnaud},
  M.~{Ashdown}, J.~{Aumont}, C.~{Baccigalupi}, A.~J. {Banday}, R.~B.
  {Barreiro}, N.~{Bartolo}, and et~al., {\it {Planck 2015 results. XIV. Dark
  energy and modified gravity}},  \href{http://arxiv.org/abs/arXiv:1502.01590}
  {{\tt {arXiv:1502.01590 [astro-ph.CO]}}}.

\end{thebibliography}\endgroup

\end{document}